\documentclass[11pt]{article}
\usepackage{epsfig}
\usepackage{amsmath}
\usepackage{amsfonts}
\usepackage{amssymb}

\numberwithin{equation}{section}

\title{The construction of Frobenius manifolds from KP tau-functions}
\author{J.W. van de Leur\footnote{JvdL
is financially supported by the
Netherlands Organization for Scientific Research (NWO).}\hskip 2cm R. Martini\\
\\
\\
Faculty of Mathematical Sciences,\\
University of Twente,\\
P.O.Box 217, 7500 AE  Enschede,\\
The Netherlands\\
fax: 31-53 489 4824\\
}

\newtheorem{lemma}{Lemma}[section]
\newtheorem{proposition}{Proposition}[section]
\newtheorem{theorem}{Theorem}[section]

\begin{document}

\maketitle

\begin{tabular}[h]{lcl}
1991 MSC &:& 22E65, 22E67, 22E70, 35Q53, 81R10, 81T40\\
Keywords &:& Frobenius manifold, WDVV equations, KP hierarchy\\
\ && tau-functions, representation theory, Grassmannian
\end{tabular}

\begin{abstract}Frobenius manifolds (solutions of WDVV equations) in canonical
coordinates are determined  by the system of Darboux--Egoroff equations.
This system of partial differential equations appears as a specific subset of
the $n$-component KP hierarchy. KP representation theory and the related
Sato infinite Grassmannian are used to construct solutions of this
Darboux--Egoroff system and the related Frobenius manifolds. Finally
we show that for these solutions Dubrovin's isomonodromy
tau-function can be expressed in the KP tau-function.
\end{abstract}

\section{Introduction}

\noindent
In the beginning of the 90's in the physics literature on two-dimensional field
theory a remarkable and amazingly rich system of partial differential
equations emerged. Roughly speaking, this system describes the conditions for a
function $F=F(t)$ of the variable $t=(t^1,t^2,\ldots,t^n)$ such that
the third-order derivatives define structure constants of an associative
algebra.
These equations are commonly known as the
Witten--Dijkgraaf--E. Verlinde--H. Verlinde (WDVV) equations \cite{W1},
\cite{DVV}.
{}From the geometric point of view the WDVV equations describe the conditions
defining
a Frobenius manifold. This concept of Frobenius manifold was introduced and
extensively studied by Dubrovin, whose lecture notes \cite{Du2} constitute
the primary reference for Frobenius manifolds and many of their applications.
The lecture notes of Manin \cite{Ma} are also a very good general reference.
Frobenius manifolds have appeared in a wide range of settings, including
quantum cohomology \cite{KM}, Gromov--Witten invariants, unfolding of
singularities, reflection groups and integrable systems. Thus Frobenius
manifolds
(WDVV equations) are relevant in describing some deep geometrical phenomena.
So it is expected that these Frobenius manifold equations are rather difficult
to solve. Suprisingly some exact explicit solutions of this system of nonlinear
equations do exist.

The WDVV equations first appeared in 2D topological field theory. It was
derived as a system of equations for so-called primary free energy.
According to an idea of Witten the procedure of  coupling to gravity should be
described in terms of an integrable hierarchy of partial differential
equations. In this context Witten--Kontsevich \cite{W2},\cite{Ko} proved that
the
partition function is a particular tau-function of the KdV hierarchy.
For general 2D topological field theories the corresponding integrable
hierarchies are not known.

The connection of Frobenius manifolds with integrable systems has been the
subject of many investigations. For instance Dubrovin (see e.g. \cite{Du2}, \S
6)
made extensive study of Frobenius manifolds in relation to semi-classical
approximations (dispersionless limit, Witham averaging) of integrable
hierarchies of partial differential equations. Here also tau-functions emerge,
but
their representation theoretical meaning remains unclear and underexposed.
Recently
tau-functions also reappear in studying one-loop approximations
\cite{DuZ},\cite{Gi}.

The particular class of semisimple Frobenius manifolds may be effectively
studied in the so-called canonical coordinates. In these coordinates
Frobenius manifolds are determined by the classical Darboux--Egoroff equations,
a system of differential equations, playing a major part in many investigations
in classical differential geometry.

It is observed that these Darboux--Egoroff equations are a special case of
the $n$-component KP hierarchy. This observation enables us to study
Frobenius manifolds in the context of the KP hierarchy. In particular
this implies that we have the machinery from the representation theory
for the KP hierarchy at our disposal and may take advantage of
it to produce solutions. This is the subject of the present paper.

The paper is devoted to the construction of Frobenius manifolds
by considering the WDVV equations in the context of the KP hierarchy
and to construct solutions in terms of appropriate classes of tau-functions
emerging in the representation theory of the KP hierarchy.

We summarize the contents of the paper. In \S 2 we explain the construction of
the
semi-infinite wedge representation of the group $GL_{\infty}$ and write down
the
condition for the $GL_{\infty}$-orbit ${\cal O}_m$ of the highest weight vector
$|m\rangle$. The resulting equation is called the KP hierarchy in the fermionic
picture. Moreover we briefly discuss the formulation within Sato's
Grassmannian.
\S 3 is devoted to bosonization of the fermionic picture.
We express the fermionic fields in terms of bosonic fields and
determine the conditions for elements of orbits ${\cal O}_m$ in bosonic terms.
Using the so-called boson-fermion correspondence we reformulate in \S 4
the KP hierarchy in the bosonic setting. Introducing formal pseudodifferential
 operators we obtain Sato's equation, another reformulation of the KP
hierarchy. In \S 5, the central part of the paper, we construct solutions of
the Darboux--Egoroff system by considering this system as a special case
of the Sato equation and applying the results described in the previous
sections and
furthermore by introducing appropriate well-chosen tau-functions. The relevance
of the orthogonal group is briefly explained.
Using the KP wave function corresponding to all solutions of \S 5, we construct
in
\S 6 specific eigenfunctions that determine the Frobenius manifold.
We find an expression for the flat coordinates and express Dubrovin's
isomonodromy
tau-function in terms of the KP tau-function.
Finally in \S 7 as an illustration we describe the simplest example in all
detail.

For notations and general background we refer to Dubrovin \cite{Du2} and
Kac and van de Leur \cite{KV}.

\section[The semi-infinite wedge
representation]{The semi-infinite wedge
representation of the group $GL_{\infty}$ and Sato's Grassmannian}

\noindent Consider the infinite complex matrix group
\[
GL_{\infty} = \{ A = (a_{ij})_{i,j \in {\mathbb Z}+\frac{1}{2}}|A\
\text{is invertible and all but a finite number of}\ a_{ij} -
\delta_{ij}\ \text{are}\ 0\}
\]
and its Lie algebra
\[
gl_{\infty} = \{ a = (a_{ij})_{i,j \in {\mathbb Z}+\frac{1}{2}}|\
\text{all but a finite number of}\ a_{ij}\ \text{are}\ 0\}
\]
with bracket $[a,b] = ab-ba$.  The Lie algebra $gl_{\infty}$ has a
basis consisting of matrices $E_{ij},\ i,j \in {\mathbb Z} + \frac{1}{2}$,
where
$E_{ij}$ is the matrix with a $1$ on the $(i,j)$-th entry and zeros
elsewhere.
Let ${\mathbb C}^{\infty} = \bigoplus_{j \in {\mathbb Z}+\frac{1}{2}} {\mathbb
C} v_{j}$ be an infinite dimensional complex vector space with fixed
basis $\{ v_{j}\}_{j \in {\mathbb Z}+\frac{1}{2}}$.  Both the group
$GL_{\infty}$ and its Lie algebra $gl_{\infty}$ act linearly on
${\mathbb C}^{\infty}$ via the usual formula:
\[
E_{ij} (v_{k}) = \delta_{jk} v_{i}.
\]

The well-known semi--infinite wedge representation is
constructed as follows \cite{KP2} (see also \cite{KR} and \cite{KV}).
The semi-infinite wedge space $F =
\Lambda^{\frac{1}{2}\infty} {\mathbb C}^{\infty}$ is the vector space
with a basis consisting of all semi-infinite monomials of the form
$v_{i_{1}} \wedge v_{i_{2}} \wedge v_{i_{3}} \ldots$, where $i_{1} >
i_{2} > i_{3} > \ldots$ and $i_{\ell +1} = i_{\ell} -1$ for $\ell >>
0$.  We can now define representations $R$ of $GL_{\infty}$ and $r$
of $gl_{\infty}$ on $F$ by
\begin{equation}
\begin{aligned}
\label{2.1}
R(A) (v_{i_{1}} \wedge v_{i_{2}} \wedge v_{i_{3}} \wedge \cdots) &= A
v_{i_{1}} \wedge Av_{i_{2}} \wedge Av_{i_{3}} \wedge \cdots ,\\
r(a) (v_{i_{1}} \wedge v_{i_{2}} \wedge v_{i_{3}} \wedge \cdots ) &=
\sum_{k} v_{i_{1}} \wedge v_{i_{2}} \wedge \cdots \wedge v_{i_{k-1}}
\wedge av_{i_{k}} \wedge v_{i_{k+1}} \wedge \cdots .
\end{aligned}
\end{equation}
These equations are related by the usual formula:
\[
\exp (r(a)) = R(\exp a)\ \text{for}\ a \in gl_{\infty}.
\]
In order to perform calculations later on, it is convenient to
introduce a larger group
\[
\begin{aligned}
\overline{GL_{\infty}} = \{ A = (a_{ij})_{i,j \in {\mathbb Z}+\frac{1}{2}}|&A\
\text{is invertible and all but a finite}\\
&\text{ number of}\ a_{ij} -
\delta_{ij}\ \text{with}\ i\ge j\ \text{are}\ 0\}
\end{aligned}
\]
and its Lie algebra
\[
\overline{gl_{\infty}} = \{ a = (a_{ij})_{i,j \in {\mathbb Z}+\frac{1}{2}}|\
\text{all but a finite number of}\ a_{ij}\ \text{with}\ i\ge j\
\text{are}\ 0\}.
\]
Both $\overline{GL_{\infty}}$ and $\overline{gl_{\infty}}$ act on a
completion $\overline{\mathbb C^\infty}$ of the space $\mathbb C^\infty$,
where
\[
\overline{\mathbb C^\infty}=\{\sum_j c_jv_j | c_j=0\ \text{for}\
j>>0\}.
\]
It is easy to see that the representations $R$ and $r$ extend to
representations of $\overline{GL_{\infty}}$ and
$\overline{gl_{\infty}}
$ on the space $F$.

The representation $r$ of $gl_{\infty}$ and $\overline{gl_{\infty}}$
can be described
in  terms of wedging and contracting
operators in $F$
(see e.g. \cite{KP2}, \cite{KR}). Let $v_j^*$ be the linear functional on
$\mathbb C^\infty$
defined by  $\langle v_i^*,v_j\rangle :=v_i^*(v_j)=\delta_{ij}$ and let
$\mathbb C^{\infty *}=\bigoplus_{j \in {\mathbb Z}+\frac{1}{2}} {\mathbb
C} v_{j}^*$ be the restricted dual of $\mathbb C^{\infty}$,
then for any $w\in \mathbb C^{\infty }$, we define  a wedging operator
$\psi^+[w]$
on $F$ by
\begin{equation}
\label{2.2}
\psi^{+}[w] (v_{i_{1}} \wedge v_{i_{2}} \wedge \cdots ) =
w\wedge v_{i_{1}} \wedge v_{i_{2}} \cdots .
\end{equation}
Let $w^*\in\mathbb C^{\infty *}$, we define a contracting operator
\begin{equation}
\label{2.3}
\psi^{-}[w^*] (v_{i_{1}} \wedge v_{i_{2}} \wedge \cdots ) =
\sum_{s=1}^\infty (-1)^{s+1}
\langle w^*,v_{i_s}\rangle v_{i_{1}} \wedge v_{i_{2}} \wedge \cdots \wedge
v_{i_{s-1}} \wedge v_{i_{s+1}} \wedge \cdots .
\end{equation}
For simplicity we write
\begin{equation}
\label{2.4}
\psi^{+}_{j}=\psi^{+}[v_{-j}],\qquad
\psi^{-}_{j}=\psi^{-}[v_j^*]\qquad\text{for }j \in {\mathbb Z} +
\frac{1}{2}
\end{equation}
These operators satisfy the following relations
$(i,j \in {\mathbb Z}+\frac{1}{2}, \lambda ,\mu = +,-)$:
\[
\psi^{\lambda}_{i} \psi^{\mu}_{j} + \psi^{\mu}_{j}
\psi^{\lambda}_{i} = \delta_{\lambda ,-\mu} \delta_{i,-j},
\]
hence they generate a Clifford algebra, which we denote by ${\cal C}\ell$.

Introduce the following elements of $F$ $(m \in {\mathbb Z})$:
\[
|m\rangle = v_{m-\frac{1}{2} } \wedge v_{m-\frac{3}{2} } \wedge
v_{m-\frac{5}{2} } \wedge \cdots .
\]
It is clear that $F$ is an irreducible ${\cal C}\ell$-module generated
by the vacuum $|0\rangle$ such that
\[
\psi^{\pm}_{j} |0\rangle = 0 \ \text{for}\ j > 0 .
\]
It is straightforward that the representation $r$ is given by the
following formula:
\begin{equation}
\label{2.5}
r(E_{ij}) = \psi^{+}_{-i} \psi^{-}_{j}.
\end{equation}
Define the {\it charge decomposition}
\begin{equation}
\label{2.6}
F = \bigoplus_{m \in {\mathbb Z}} F^{(m)}
\end{equation}
by letting
\begin{equation}
\label{2.7}
\text{charge}\ |0\rangle  = 0\ \text{and charge } \psi^{\pm}_{j} =
\pm 1.
\end{equation}
It is clear that the charge decomposition is invariant with respect
to $r(g\ell_{\infty})$ (and hence with respect to $R(GL_{\infty})$).
Moreover, it is easy to see that each $F^{(m)}$ is irreducible with
respect to $g\ell_{\infty}$ (and $GL_{\infty}$).  Note that
$|m\rangle$ is its highest weight vector, i.e.
\[
\begin{aligned}
\ &r(E_{ij})|m\rangle = 0 \ \text{for}\ i < j, \\
\ &r(E_{ii})|m\rangle = 0\  (\text{resp.}\ = |m\rangle ) \ \text{if}\ i > m\
(\text{resp. if}\ i < m).
\end{aligned}
\]
Let $w\in F$, we define the Annihilator space  $Ann (w)$ of $w$
as follows:
\begin{equation}
\label{2.8}
Ann(w)=\{v\in\mathbb C^\infty | v\wedge w=0\}.
\end{equation}
Notice that $Ann(w)\ne 0$, since $v_j\in Ann(w)$ for $j<<0$.
This Annihilator space for perfect (semi--infinite) wedges
$w\in F^{(m)}$ is related to the
$GL_{\infty}\text{-orbit}$
\[
{\cal O}_m
= R(GL_{\infty})|m\rangle \subset F^{(m)}
\]
of the highest weight vector $|m\rangle$ as follows.
Let $A=(A_{ij})_{i,j \in {\mathbb Z}}\in GL_\infty$, denote by
$A_j=\sum_{i\in{\mathbb Z}} A_{ij}v_i$  then by (2.8)
\begin{equation}
\label{2.9}
\tau_m=R(A)|m\rangle =A_{m-{1\over 2}}\wedge
A_{m-{3\over 2}}\wedge A_{m-{5\over 2}}\wedge
\cdots,
\end{equation}
with $A_{-j}=v_{-j}$ for $j>>0$.
Notice that since $\tau_m$ is a perfect
(semi-infinite) wedge
\[
Ann(\tau_m)=\sum_{j<m} \mathbb CA_j
\subset \mathbb C^\infty.
\]
The following theorem also characterizes the group orbit for a proof
see \cite{KP2}\cite{KR}:
\begin{theorem}
\label{t1}
Let $\tau_m\in F^{(m)}$, then $\tau_m\in {\cal O}_m$ if and only if
$\tau_m$ satisfies the (fermionic) KP hierarchy:
\begin{equation}
\label{1.3.1}
\sum_{k \in {\mathbb Z}+\frac{1}{2}} \psi^{+}_{k} \tau_m \otimes \psi^{-}_{-k}
\tau_m = 0.
\end{equation}
\end{theorem}
It is obvious from the construction that if $w\in \mathbb{C}^\infty$
and $\tau_m\in {\cal O}_m$ that $w\wedge \tau_m\in {\cal O}_{m+1}$.
In fact one has the following usefull lemma.
\begin{lemma}
\label{l1}
Let $\tau_m\in {\cal O}_m$, $w\in \mathbb{C}^\infty$ and $w^*\in
\mathbb{C}^{\infty *}$. If $\psi^+[w]\tau_m\ne 0$
(resp. $\psi^-[w^*]\tau_m\ne 0$), then   $\psi^+[w]\tau_m\in {\cal
O}_{m+1}$
(resp. $\psi^-[w^*]\tau_m\in {\cal
O}_{m-1}$).
\end{lemma}
{\bf Proof}. We only have to prove the statement for
$\psi^-[w^*]\tau_m$. Let $\psi^-[w^*]\otimes \psi^-[w^*]$ act on
(\ref{1.3.1}), then we obtain
\[
\begin{aligned}
0&=\sum_{k \in {\mathbb Z}+\frac{1}{2}} \psi^{+}_{k}\psi^-[w^*] \tau_m \otimes
\psi^{-}_{-k}\psi^-[w^*]
\tau_m
-\sum_{k \in {\mathbb Z}+\frac{1}{2}}\tau_m \otimes \langle w^*,
v_{-k}\rangle\psi^{-}_{-k}\psi^-[w^*]
\tau_m\\
 &=\sum_{k \in {\mathbb Z}+\frac{1}{2}} \psi^{+}_{k}\psi^-[w^*] \tau_m \otimes
\psi^{-}_{-k}\psi^-[w^*]
\tau_m-
\tau_m \otimes\psi^-[w^*]\psi^-[w^*]
\tau_m\\
\end{aligned}
\]
Since the last term is clearly zero we obtain the desired result.
\hfill$\square$

Choose a positive integer $n$ and relabel the basis vectors $v_i$ as
follows. Define
for $j \in {\mathbb Z},\ 1 \leq j \leq n,\ k \in
{\mathbb Z} + \frac{1}{2}$:
\begin{equation}
\label{2.10}
v^{(j)}_{k} = v_{nk - \frac{1}{2}(n-2j+1)},
\end{equation}
and identify
\begin{equation}
\label{2.11}
v^{(j)}_{k} = t^{-k-\frac{1}{2}}e_j,
\end{equation}
where $e_j$, $1\le j\le n$, is abasis of $\mathbb{C}^n$. We can thus
write the vectors $A_\ell$ in (\ref{2.9}) as
\begin{equation}
\label{2.12}
A_\ell=A_\ell(t)=\sum_{j=1}^n\left (\sum_{i\in\mathbb Z+{1\over 2}}
A_{ni- \frac{1}{2}(n-2j+1),\ell}t^{-i-{1\over 2}}\right )e_j,
\end{equation}
hence as a vector in $H=(\mathbb{C}[t,t^{-1}])^n$.
In this way we can identify
$Ann(\tau_m)$ with a subspace
$W_{\tau_m}=\sum_{j<m}\mathbb C A_j(t)$ of the
space $H$ and hence with a point in an infinite (polynomial)
Grassmannian $Gr$. A point of $Gr$  is a linear subspaces
of $H$ which contains
\[
H_\ell:=\sum_{j=1}^n\sum_{i=\ell}^\infty \mathbb C t^ie_j
\]
for $\ell>>0$. Now $Gr=\cup_{m\in\mathbb Z}Gr_m$ (disjoint union) with
\[
Gr_m=\{W\in Gr| H_\ell\subset W\ \text{and}\ \dim W/H_\ell=\ell n+m
\ \text{for}\ \ell>>0\},
\]
and we can construct a cannonical map
\[
\phi :{\cal O}_m\to Gr_m,\qquad \phi(\tau_m)
=W_{\tau_m}:=\sum_{i<m} \mathbb C A_i(t).
\]
It is clear that $\phi (|mn\rangle )=H_{-m}$ and that $\phi$ is
surjective with fibers $\mathbb C^\times$. This construction
is due to Sato [S].

\section{The boson-fermion correspondence}

\noindent
The relabeling of the $v_i$'s given by (\ref{2.10}) induces a
relabeling of the $\psi^{\pm}_j$'s, viz.,

\[
\psi^{\pm (j)}_{k} = \psi^{\pm}_{nk \pm \frac{1}{2}
(n-2j+1)}.
\]
Notice that with this relabeling we have:
\[
\psi^{\pm (j)}_{k}|0\rangle = 0\ \text{for}\ k > 0.
\]
Besides the charge decomposition, we also introduce  an {\it energy
decomposition}
defined by
\begin{equation}
\label{2.1.2}
\text{energy}\ |0\rangle = 0,\ \text{energy}\ \psi^{\pm (j)}_{k} =
-k.
\end{equation}
Note that energy on $F$ is never negative.
Introduce the  fermionic fields $(z \in {\mathbb C}^{\times})$ by
\begin{equation}
\label{2.1.4}
\psi^{\pm (j)}(z) = \sum_{k \in {\mathbb
Z}+\frac{1}{2}} \psi^{\pm
(j)}_{k} z^{-k-\frac{1}{2}},
\end{equation}
and bosonic fields $(1 \leq i,j \leq n)$ by
\begin{equation}
\label{2.1.5}
\alpha^{(ij)}(z)= \sum_{k \in {\Bbb Z}} \alpha^{(ij)}_{k} z^{-k-1}
= :\psi^{+(i)}(z) \psi^{-(j)}(z):
\end{equation}
where $:\ :$ stands for the {\it normal ordered product} defined in
the usual way $(\lambda ,\mu = +$ or $-$):
\begin{equation}
\label{2.1.6}
:\psi^{\lambda (i)}_{k} \psi^{\mu (j)}_{\ell}: = \begin{cases} \psi^{
\lambda (i)}_{k}
\psi^{\mu (j)}_{\ell}\ &\text{if}\ \ell \ge k ,\\
-\psi^{\mu (j)}_{\ell} \psi^{\lambda (i)}_{k} &\text{if}\ \ell <k
.\end{cases}
\end{equation}
One checks (using e.g. the Wick formula) that the operators
$\alpha^{(ij)}_{k}$ satisfy the commutation relations of the affine
algebra $gl_{n}({\mathbb C})^{\wedge}$ with central charge $1$, i.e.:
\begin{equation}
\label{2.1.7}
[\alpha^{(ij)}_{p},\alpha^{(k\ell)}_{q}] =
\delta_{jk}\alpha^{(i\ell )}_{p+q}
- \delta_{i\ell} \alpha^{(kj)}_{p+q} + p\delta_{i\ell}
\delta_{jk}\delta_{p,-q},
\end{equation}
and that
\begin{equation}
\label{2.1.8}
\alpha^{(ij)}_{k}|m\rangle = 0 \ \text{if}\ k > 0 \ \text{or}\ k = 0\
\text{and}\ i < j.
\end{equation}
The operators $\alpha^{(i)}_{k} \equiv \alpha^{(ii)}_{k}$
satsify the canonical commutation relation of the associative
oscillator algebra,  which we
denote by ${\cal a}$:
\begin{equation}
\label{2.1.9}
[\alpha^{(i)}_{k},\alpha^{(j)}_{\ell}] =
k\delta_{ij}\delta_{k,-\ell},
\end{equation}
and one has
\begin{equation}
\label{2.1.10}
\alpha^{(i)}_{k}|m\rangle = 0 \ \text{for}\ k > 0.
\end{equation}

It is easy to see that restricted to $g\ell_{n}({\mathbb C})^{\wedge}$,
$F^{(0)}$ is its basic highest weight representation (see \cite{K}).

In order to express the fermionic fields $\psi^{\pm (i)}(z)$ in terms of
the bosonic fields $\alpha^{(i)}(z)$, we need some additional operators
$Q_{i},\ i = 1,\ldots ,n$, on $F$.  These operators are uniquely defined by
the following conditions:
\begin{equation}
\label{2.1.12}
Q_{i}|0\rangle = \psi^{+(i)}_{-\frac{1}{2}} |0\rangle ,\ Q_{i}\psi^{\pm
(j)}_{k} = (-1)^{\delta_{ij}+1} \psi^{\pm
(j)}_{k\mp \delta_{ij}}Q_{i}.
\end{equation}
They satisfy the following commutation relations:
\begin{equation}
\label{2.1.13}
Q_{i}Q_{j} = -Q_{j}Q_{i}\ \text{if}\ i \neq j,\ [\alpha^{(i)}_{k},Q_{j}] =
\delta_{ij} \delta_{k0}Q_{j}.
\end{equation}
\begin{theorem}
\label{t2.1}  (\cite{DJKM1}, \cite{JM})
\begin{equation}
\label{2.1.14}
\psi^{\pm (i)}(z) = Q^{\pm 1}_{i}z^{\pm \alpha^{(i)}_{0}} \exp
(\mp \sum_{k < 0} \frac{1}{k} \alpha^{(i)}_{k}z^{-k})\exp(\mp
\sum_{k > 0} \frac{1}{k} \alpha^{(i)}_{k} z^{-k}).
\end{equation}
\end{theorem}
{\bf Proof}. See \cite{TV}.

The operators on the right-hand side of (\ref{2.1.14}) are called vertex
operators.  They made their first appearance in string theory (cf.
\cite{FK}).

We can describe now the $n$-component boson-fermion
correspondence.  Let ${\mathbb C}[x]$ be the space of polynomials in
indeterminates $x = \{ x^{(i)}_{k}\},\ k = 1,2,\ldots ,\ i =
1,2,\ldots ,n$.  Let $L$ be a lattice with a basis $\delta_{1},\ldots
,\delta_{n}$ over ${\mathbb Z}$ and the symmetric bilinear form
$(\delta_{i}|\delta_{j}) = \delta_{ij}$, where $\delta_{ij}$ is the
Kronecker symbol.  Let
\begin{equation}
\label{2.2.1}
\varepsilon_{ij} = \begin{cases} -1 &\text{if $i > j$} \\
1 &\text{if $i \leq j$.} \end{cases}
\end{equation}
Define a bimultiplicative function $\varepsilon :\ L \times L\to \{
\pm 1 \}$ by letting
\begin{equation}
\label{2.2.2}
\varepsilon (\delta_{i}, \delta_{j}) = \varepsilon_{ij}.
\end{equation}
Let $\delta = \delta_{1} + \ldots + \delta_{n},\ M = \{ \gamma \in
L|\ (\delta | \gamma ) = 0\}$, $\Delta = \{ \alpha_{ij} :=
\delta_{i}-\delta_{j}| i,j = 1,\ldots ,n,\ i \neq j \}$.  Of course
$M$ is the root lattice of $s\ell_{n}({\mathbb C})$, the set $\Delta$
being the root system.

Consider the vector space ${\mathbb C}[L]$ with basis $e^{\gamma}$,\
$\gamma \in L$, and the following twisted group algebra product:
\begin{equation}
\label{2.2.3}
e^{\alpha}e^{\beta} = \varepsilon (\alpha ,\beta)e^{\alpha +
\beta}.
\end{equation}
Let $B = {\mathbb C}[x] \otimes_{\mathbb C} {\mathbb C}[L]$ be the tensor
product of algebras.  Then the $n$-component boson-fermion
correspondence is the vector space isomorphism
\begin{equation}
\label{2.2.4}
\sigma :F \to B,
\end{equation}
given by
\begin{equation}
\label{2.2.5}
\sigma (\alpha^{(i_{1})}_{-m_{1}} \ldots
\alpha^{(i_{s})}_{-m_{s}}Q_1^{k_{1}}\ldots Q_n^{k_{n}}|0\rangle ) = m_{1}
\ldots
m_{s}x^{(i_{1})}_{m_{1}} \ldots x^{(i_{s})}_{m_{s}} \otimes
e^{k_{1}\delta_{1} + \ldots + k_{n}\delta_{n}} .
\end{equation}
The transported charge and energy then will be as follows:
\begin{equation}
\label{2.2.6}
\begin{aligned}
\ &\text{charge }p(x)\otimes e^{\gamma} = (\delta |\gamma),\\
\ &\text{energy }x^{(i_{1})}_{m_{1}} \ldots x^{(i_{s})}_{m_{s}} \otimes
e^{\gamma} = m_{1} + \ldots + m_{s} + { \frac{1}{2}} (\gamma |\gamma).
\end{aligned}
\end{equation}
We denote the transported charge decomposition by
\[B = \bigoplus_{m \in {\mathbb Z}} B^{(m)}.
\]

The transported action of the operators $\alpha^{(i)}_{m}$ and $Q_{j}$ looks
as follows:
\begin{equation}\label{2.2.8}
\begin{cases}
\sigma \alpha^{(j)}_{-m}\sigma^{-1}(p(x) \otimes e^{\gamma}) =
mx^{(j)}_{m}p(x)\otimes e^{\gamma},\ \text{if}\ m > 0, &\  \\
\sigma \alpha^{(j)}_{m} \sigma^{-1}(p(x) \otimes e^{\gamma}) = \frac{\partial
p(x)}{\partial x_{m}} \otimes e^{\gamma},\ \text{if}\ m > 0, &\  \\
\sigma \alpha^{(j)}_{0} \sigma^{-1} (p(x) \otimes e^{\gamma}) =
(\delta_{j}|\gamma ) p(x) \otimes e^{\gamma} , &\ \\
\sigma Q_{j} \sigma^{-1} (p(x) \otimes e^{\gamma}) = \varepsilon
(\delta_{j},\gamma)  p(x) \otimes e^{\gamma + \delta_{j}}
. & \
\end{cases}
\end{equation}
The transported action of the fermionic fields is as follows:
\begin{equation}
\label{2.2.9}
\sigma\psi^{\pm(j)}(z)\sigma^{-1}=
e^{\pm \delta_j}z^{\pm \delta_j}\exp (\pm \sum^{\infty}_{k=1}
x^{(j)}_{k} )
\exp
( \mp \sum^{\infty}_{k=1}  \frac{\partial}{\partial
x^{(j)}_{k}} \frac{z^{-k}}{k})
\end{equation}

We will now determine the second part of the boson--fermion
correspondence, i.e., we want to determine $\sigma(\tau_m)$, where
$\tau_m$ is given by (\ref{2.9}).
Since all spaces $F^{(m)}$ give a similar representation of
$gl_\infty$, we will restrict our attention to the case that $m=0$ and we
write $\tau$ instead of $\tau_0$. We will generalize the proof of
Theorem 6.1 of \cite{KR}. For this purpose we have to introduce
elements $\Lambda^{(j)}_\ell\in\overline{gl_\infty}$, $1\le j\le n$,
$\ell\in\mathbb{N}$, by
\begin{equation}
\label{3.1}
\Lambda^{(j)}_\ell=
\sum_{k\in \mathbb{Z}+\frac{1}{2}}
E_{nk-\frac{1}{2}(n-2j+1),n{k+\ell}-\frac{1}{2}(n-2j+1)}.
\end{equation}
Notice that $\Lambda^{(j)}_\ell=(\Lambda^{(j)}_1)^\ell$,
$r(\Lambda^{(j)}_\ell)=\alpha_\ell^{(j)}$
 and that
$\exp \Lambda^{(j)}_\ell\in \overline{Gl_\infty}$.
With the relabeling $|0\rangle$ becomes
\[
|0\rangle=
v_{-\frac{1}{2}}^{(n)}\wedge
v_{-\frac{1}{2}}^{(n-1)}\wedge\cdots\wedge
v_{-\frac{1}{2}}^{(1)}\wedge
v_{-\frac{3}{2}}^{(n)}\wedge
v_{-\frac{3}{2}}^{(n-1)}\wedge\cdots,
\]
and
\[
Q_iQ_j^{-1}|0\rangle=
(-)^{n-j}v_{\frac{1}{2}}^{(i)}v_{-\frac{1}{2}}^{(n)}\wedge
v_{-\frac{1}{2}}^{(n-1)}\wedge\cdots\wedge
v_{-\frac{1}{2}}^{(j+1)}\wedge
v_{-\frac{1}{2}}^{(j-1)}\wedge\cdots\wedge
v_{-\frac{1}{2}}^{(1)}\wedge
v_{-\frac{3}{2}}^{(n)}\wedge
v_{-\frac{3}{2}}^{(n-1)}\wedge\cdots.
\]
We now want to determine $\sigma(\tau)$, where
\begin{equation}
\label{3.2}
\tau=R(A)|0\rangle=A_{-\frac{1}{2}}\wedge
A_{-\frac{3}{2}}\wedge
A_{-\frac{5}{2}}\wedge
\cdots,\ \text{with } A_{-p}=v_{-p} \ \text{for all } p>P>>0.
\end{equation}
Let $\sigma(\tau)= \sum_{\alpha\in M}\tau_\alpha(x)e^\alpha$, we want
to compute
\[
\sigma\left ( R\left ( \exp \left (
\sum_{j=1}^n\sum_{k=1}^\infty y_k^{(j)}\Lambda^{(j)}_k
\right )\right )\tau\right )=\exp \left (\sum_{j=1}^n\sum_{k=1}^\infty
y_k^{(j)}
\frac{\partial}{\partial x_k^{(j)}}\right )
\sum_{\alpha\in M}\tau_\alpha(x)e^\alpha.
\]
Now let $F_\alpha(y)$ denote the coefficient of $1 \otimes e^\alpha$ in this
expression, then
\[
F_\alpha(y)=
\exp \left (\sum_{j=1}^n\sum_{k=1}^\infty
y_k^{(j)}
\frac{\partial}{\partial x_k^{(j)}}\right )
\tau_\alpha(x)|_{x=0}=\tau_\alpha(x+y)|_{x=0}=\tau_\alpha(y).
\]
So $\tau_\alpha(y)$ is the coefficient of $1\otimes e^\alpha$ in
\[
\sigma\left ( R\left ( \exp \left (
\sum_{j=1}^n\sum_{k=1}^\infty y_k^{(j)}\Lambda^{(j)}_k
\right )A\right )|0\rangle\right ).
\]
Now let $\alpha=\sum_{j=1}^nk_j\delta_j$ then
\[
1\otimes e^\alpha=\sigma(Q_1^{k_1}Q_2^{k_2}\cdots Q_n^{k_n}|0\rangle ),
\]
hence $\tau_\alpha(y)$ is the coefficient of
\[
\begin{aligned}
R&\left ( \exp \left (
\sum_{j=1}^n\sum_{k=1}^\infty y_k^{(j)}\Lambda^{(j)}_k
\right )A\right )|0\rangle
=R\left (\left (\sum_{j=1}^n\sum_{k=0}^\infty
S_k(y^{(j)})\Lambda^{(j)}_k\right )A\right )|0\rangle\\
\ &
=R\left ( \sum_{\ell<0} \sum_{j=1}^n\sum_{q\in\mathbb{Z}+\frac{1}{2}}
\left (\sum_{k=0}^\infty
A_{n(q+k)-\frac{1}{2}
(n-2j+1),n(q)-\frac{1}{2}
(n-2j+1)}S_k(y^{(j)})\right )
E_{nq-\frac{1}{2}
(n-2j+1),\ell}\right )|0\rangle,
\end{aligned}
\]
where $S_k(y)$ are the elementary Schur functions defined by
$\sum_{k\in\mathbb{Z}}S_k(y)z^k=\exp (\sum_{k=1}^\infty y_kz^k)$.
Using formula (4.48) of \cite{KR}, i.e.,
\[
R(A)|0\rangle=
\sum_{j_{-\frac{1}{2}}>
j_{-\frac{3}{2}}>
j_{-\frac{5}{2}}>\cdots}\det \left (
A_{j_{-\frac{1}{2}},
j_{-\frac{3}{2}},
j_{-\frac{5}{2}},\cdots}^{{-\frac{1}{2}},
{-\frac{3}{2}},
{-\frac{5}{2}},\cdots}\right )v_{j_{-\frac{1}{2}}}\wedge
v_{j_{-\frac{3}{2}}}\wedge
v_{j_{-\frac{5}{2}}}\wedge\cdots,
\]
where
$
A_{j_{-\frac{1}{2}},
j_{-\frac{3}{2}},
j_{-\frac{5}{2}},\cdots}^{{-\frac{1}{2}},
{-\frac{3}{2}},
{-\frac{5}{2}},\cdots}$ denotes the matrix located at the intersection
of the rows $j_{-\frac{1}{2}},
j_{-\frac{3}{2}},
j_{-\frac{5}{2}},\cdots$ and the columns ${-\frac{1}{2}},
{-\frac{3}{2}},
{-\frac{5}{2}},\cdots$ of the matrix $A$, we can calculate
$\tau_\alpha(y)$ if we can determine $Q_1^{k_1}Q_2^{k_2}\cdots
Q_n^{k_n}|0\rangle$
as a perfect simple wedge. This is in general quite complicated, so we
assume for the moment that
\[
Q_1^{k_1}Q_2^{k_2}\cdots
Q_n^{k_n}|0\rangle=\lambda_\alpha v_{j_{-\frac{1}{2}}}\wedge
v_{j_{-\frac{3}{2}}}\wedge
v_{j_{-\frac{5}{2}}}\wedge\cdots,
\]
with $j_{-q}=-q$ for all $q>Q>>0$ and $\lambda_\alpha=\pm 1$,
then
\[
\tau_\alpha(y)=\lambda_\alpha
\det\left(
\sum_{\ell<0}
\sum_{r=j_{-\frac{1}{2}},
j_{-\frac{3}{2}},
j_{-\frac{5}{2}},\cdots}
\sum_{
{1\le j\le n, q\in\mathbb{Z}+\frac{1}{2}}\atop
{
nq-\frac{1}{2}(n-2j+1)=r}
}
\left (\sum_{k=0}^\infty A_{r+nk,\ell}S_k(y^{(j)})\right )E_{r,\ell}\right
).
\]
Finally notice that this is in fact only a finite determinant of size
$R=\max (P,Q)$, hence we have determined
\begin{proposition}
\label{p1}
Let $A=(A_{i,j})_{i,j\in\mathbb{Z}+\frac{1}{2}}\in GL_\infty$ be such
that $A_{ij}=\delta_{ij}$ for $j<-P$ then
$\sigma(R(A)|0\rangle=\sum_{\alpha\in M}\tau_\alpha(x)e^\alpha$.
Assume that $\alpha=\sum_{j=1}^nk_j\delta_j$ and suppose that
\[
Q_1^{k_1}Q_2^{k_2}\cdots
Q_n^{k_n}|0\rangle=\lambda_\alpha v_{j_{-\frac{1}{2}}}\wedge
v_{j_{-\frac{3}{2}}}\wedge
v_{j_{-\frac{5}{2}}}\wedge\cdots,
\]
with
$j_{-\frac{1}{2}}>
j_{-\frac{3}{2}}>
j_{-\frac{5}{2}}\cdots$ and  $j_{-q}=-q$ for all $q>Q>>0$ and
$\lambda_\alpha=\pm 1$,
then
\[
\tau_\alpha(x)=\lambda_\alpha
\det\left(
\sum_{-R<\ell<0}
\sum_{r=j_{-\frac{1}{2}},
j_{-\frac{3}{2}}
,\cdots,
j_{-R+\frac{1}{2}}}
\sum_{
{1\le j\le n, q\in\mathbb{Z}+\frac{1}{2}}\atop
{
nq-\frac{1}{2}(n-2j+1)=r}
}
\left (\sum_{k=0}^\infty A_{r+nk,\ell}S_k(x^{(j)})\right )E_{r,\ell}\right
),
\]
where $R=\max (P,Q)$.
In particular if $1\le i<j\le n$ and $\alpha=0$, $\delta_i-\delta_j$,
$\delta_j-\delta_i$, respectively, then
$\lambda_0=1$, $\lambda_{\delta_i-\delta_j}=(-1)^{n-j}$,
$\lambda_{\delta_j-\delta_i}=(-1)^{n-i+1}$ and
$(j_{-\frac{1}{2}},
j_{-\frac{3}{2}}
,\cdots)=(-\frac{1}{2},-\frac{3}{2},-\frac{5}{2},\ldots)$,
$=(i-\frac{1}{2},-\frac{1}{2},-\frac{3}{2},\ldots,
j-n+\frac{1}{2},j-n+\frac{3}{2}\ldots)$,
$=(j-\frac{1}{2},-\frac{1}{2},-\frac{3}{2},\ldots,
i-n+\frac{1}{2},i-n+\frac{3}{2}\ldots)$, respectively.
\end{proposition}

\section{The KP hierarchy as a dynamical system}

\noindent
Using the isomorphism $\sigma$ we can reformulate the KP hierarchy
(\ref{1.3.1}) in the bosonic picture.
We start by observing that (\ref{1.3.1}) can be rewritten as follows:
\begin{equation}
\label{2.3.1}
\text{Res}_{z=0} \sum^{n}_{j=1} \psi^{+(j)}(z)\tau
\otimes \psi^{-(j)}(z)\tau  = 0,\ \tau \in F^{(0)}.
\end{equation}
Here and further $\text{Res}_{z=0} \sum_{j} f_{j}z^{j}$ (where
$f_{j}$ are independent of $z$) stands for $f_{-1}$.
Notice that for $\tau \in F^{(0)},\ \sigma (\tau) = \sum_{\gamma \in M}
\tau_{\gamma}(x)e^{\gamma}$.
 Here and further  we write $\tau_{\gamma}(x)e^{\gamma}$ for
$\tau_{\gamma} \otimes
e^{\gamma}$.  Using Theorem \ref{t2.1}, equation (\ref{2.3.1}) turns under
$\sigma
\otimes \sigma :\ F \otimes F \longrightarrow
{\mathbb C}[x^{\prime},x^{\prime \prime}]
\otimes ({\mathbb C}[L^{\prime}] \otimes {\mathbb C}[L^{\prime \prime}])$ into
the
following equations, which we call the $n$-component KP hierarchy.
Let $1 \leq a,b \leq n$, $\alpha,\ \beta\in M$:
\begin{equation}
\label{3.3.1}
\begin{aligned}
\ &\text{Res}_{z=0}( \sum^{n}_{j=1} \varepsilon (\delta_{j}, \alpha +
\delta_{a} - \beta + \delta_{b})
 z^{(\delta_{j}|\alpha + \delta_{a} -
\beta + \delta_{b} -2\delta_{j} )} \\
\ &\times \exp (\sum^{\infty}_{k=1} (x^{(j)^{\prime}}_{k} - x^{(j)^{\prime
\prime}}_{k})z^{k}) \exp(-\sum^{\infty}_{k=1}
(\frac{\partial}{\partial
x^{(j)^{\prime}}_{k}}-\frac{\partial}{\partial
x^{(j)^{\prime\prime}}_{k}}) \frac{z^{-k}}{k}) \\
\ &\tau_{\alpha + \alpha_{a_{j}}} ( x^{\prime})
\tau_{\beta - \alpha_{b_{j}}}(x^{\prime\prime}))
 =
0\ \ \ \ (\alpha ,\beta \in M).\\
\end{aligned}
\end{equation}
Define the support of $\tau$  by
$\text{supp }\tau=\{ \alpha\in M|\tau_\alpha\ne 0\}$, then for  each $\alpha
\in \ \text{supp}\ \tau$ we define the (matrix
valued) wave functions
\begin{equation}
\label{3.3.2}
V^{\pm} (\alpha,x,z) = (V^{\pm}_{ij}(\alpha ,x,z))^{n}_{i,j=1}
\end{equation}
as follows:
\begin{equation}
\label{3.3.3}
\begin{aligned}
\ &V^{\pm}_{ij}(\alpha ,x,z) :=
\varepsilon (\delta_{j} , \alpha + \delta_{i})
 z^{(\delta_{j}|\pm \alpha + \alpha_{ij})} \\
\ & \times \exp (\pm \sum^{\infty}_{k=1} x^{(j)}_{k} z^{k})
\exp(\mp \sum^{\infty}_{k=1} \frac{\partial}{\partial
x^{(j)}_{k}} \frac{z^{-k}}{k}) \tau_{\alpha  \pm
\alpha_{ij}}  (x)/\tau_{\alpha}(x)
\end{aligned}
\end{equation}
It is easy to see that equation (\ref{3.3.1}) is equivalent to the
following bilinear identity:
\begin{equation}
\label{3.3.4}
\text{Res}_{z=0}V^{+}(\alpha,x,z)\ ^{t}V^{-}(\beta, x^{\prime},z)
= 0\
\text{for all}\ \alpha ,\beta \in M,
\end{equation}
where$\ ^tV$ stands for the transposed of the matrix $V$.
Define $n \times n$ matrices $W^{\pm (m)} (\alpha ,x)$ by the
following generating series (cf. (\ref{3.3.3})):
\begin{equation}
\label{3.3.5}
\sum^{\infty}_{m=0}
W^{\pm (m)}_{ij} (\alpha ,x)(\pm z)^{-m}
= \varepsilon_{ji}z^{\delta_{ij}-1} (\exp \mp
\sum^{\infty}_{k=1} \frac{\partial}{\partial x^{(j)}_{k}}\frac{z^{-k}}{k})
\tau_{\alpha \pm
\alpha_{ij}} (x))/\tau_{\alpha} (x).
\end{equation}
Note that
\begin{equation}
\label{3.3.6}
W^{\pm (0)}(\alpha ,x) = I_{n},
\end{equation}
\begin{equation}
\label{3.3.7}
W^{\pm (1)}_{ij}(\alpha ,x) =
\begin{cases} \varepsilon_{ji}
\tau_{\alpha \pm \alpha_{ij}}/\tau_{\alpha} &\text{if}\ i \neq j \\
- \tau^{-1}_{\alpha} \frac{\partial \tau_{\alpha}}{\partial
x^{(i)}_{1}} &\text{if}\ i = j, \end{cases}
\end{equation}

We see from (\ref{3.3.3}) that $V^{\pm}(\alpha ,x,z)$ can be written in the
following form:
\begin{equation}
\label{3.3.9}
V^{\pm}(\alpha ,x,z) = \sum^{\infty}_{m=0}
W^{\pm (m)}(\alpha ,x)(\pm
z)^{-m}R^{\pm}(\alpha ,\pm z)S^\pm(x,z),
\end{equation}
where
\begin{equation}
\label{3.3.10}
\begin{aligned}
R^{\pm}(\alpha ,z) &= \sum^{n}_{i=1}
\varepsilon (\delta_{i}, \alpha ) E_{ii} (\pm z)^{\pm
(\delta_{i}|\alpha )},\\
S^{\pm}(x,z) &= \sum^{n}_{i=1} e^{\pm\sum_{j=1}^\infty  x_j^{(i)}
z^j}E_{ii}.
\end{aligned}
\end{equation}
Here  $E_{ij}$ stands for the $n \times n$ matrix whose
$(i,j)$ entry is $1$ and all other entries are zero.  Now let
$\partial=\sum_{j=1}^n \frac{\partial}{\partial x_1^{(j)}}$, then
$V^{\pm}(\alpha ,x,z)$ can be
written in
terms of formal pseudo-differential operators (see \cite{KV} for more details).
\begin{equation}
\label{3.3.11}
P^{\pm}(\alpha ) \equiv P^{\pm} (\alpha ,x,\partial ) =
I_{n} + \sum^{\infty}_{m=1} W^{\pm (m)} (\alpha ,x)\partial^{-m},\
R^{\pm}(\alpha ) = R^{\pm}(\alpha ,\partial)
\end{equation}
as follows:
\begin{equation}
\label{3.3.12}
V^{\pm}(\alpha ,x,z)
=P^{\pm } (\alpha )
R^{\pm}(\alpha)S^\pm (x,z)
\end{equation}
Since obviously
\begin{equation}
\label{3.3.13}
R^{-}(\alpha ,\partial)^{-1} = R^{+}(\alpha
,\partial)^{*},
\end{equation}
where $P^*=\sum_k (-\partial)^k\,^tP^{(k)}$ stands for the formal adjoint of
$P=\sum_kP^{(k)}\partial^k$.
Moreover one can deduce (see \cite{KV}) from the bilinear identity
(\ref{3.3.4}):
\begin{equation}
\label{3.3.14}
(P^{+}(\alpha,x,\partial)R^{+}(\alpha
-\beta,\partial)
P^{-}(\beta,x^\prime\partial)^{*})_{-} = 0
\end{equation}
for any $\alpha ,\beta \in \text{supp}\ \tau$. Here $Q_-=Q-Q_+$, where
$Q_+$ stands for the differential operator part of $Q$.

Furthermore, put $x=x^\prime$, then one deduces from (\ref{3.3.14}) with
$\alpha = \beta$ that
\begin{equation}\label{3.3.15}
P^{-}(\alpha  ) = (P^{+}(\alpha  )^{*})^{-1} ,
\end{equation}
since $R^{\pm}(0) = I_{n}$ and $P^{\pm}(\alpha) \in I_{n} +
\text{lower order terms}$.
With all these ingredients one can prove the following Lemma:

\begin{proposition}
\label{l3.4}  Let $\alpha,\beta\in \text{supp }\tau$, then
$P^{+}(\alpha)$ satisfies the Sato equations:
\begin{equation}
\label{3.4.2}
\frac{\partial P^+(\alpha)}{\partial x^{(j)}_{k}} = -(P^+(\alpha)E_{jj}
\partial^{k}  P^+(\alpha)^{-1})_{-}  P^+(\alpha)
\end{equation}
and $P^+(\alpha), \ P^+(\beta)$ satisfy
\begin{equation}
\label{3.3.16}
(P^{+}(\alpha)R^{+}(\alpha - \beta)P^{+}(\beta)^{-1})_{-} = 0\
\text{for all}\ \alpha ,\beta \in \text{supp}\ \tau .
\end{equation}
\end{proposition}
This is another formulation of the $n$-component KP hierarchy (see \cite{KV}).
Introduce the following formal pseudo-differential
operators $L(\alpha),\  C^{(j)}(\alpha )$:
\begin{equation}
\label{3.4.1}
\begin{aligned}
L(\alpha ) \equiv L(\alpha,x,\partial)
  & = P^{+}(\alpha)  \partial  P^{+}(\alpha)^{-1}, \\
C^{(j)}(\alpha) \equiv C^{(j)}(\alpha ,x,\partial) &=
P^{+}(\alpha)E_{jj} P^{+}(\alpha)^{-1}, \\
\end{aligned}
\end{equation}
then related to the Sato equation is the following linear system
\begin{equation}
\label{3.5.5}
\begin{aligned}
L(\alpha)V^{+}(\alpha ,x,z) &= zV^{+}(\alpha ,x,z) ,\\
C^{(i)}(\alpha)V^{+}(\alpha ,x,z) &=V^{+}(\alpha ,x,z) E_{ii} , \\
\frac{\partial V^{+}(\alpha ,x,z) }{\partial x^{(i)}_{k}}
&=(L(\alpha)^kC^{(i)}(\alpha))_+ V^{+}(\alpha ,x,z) .\\
\end{aligned}
\end{equation}

To end this section we write down explicitly some of the Sato
equations (\ref{3.4.2}) on the matrix elements $W^{(s)}_{ij}$ of the
coefficients $W^{(s)}(x)$ of the pseudo-differential operator
\[
P =P^+(\alpha)
= I_{n} + \sum^{\infty}_{m=1} W^{(m)}(x) \partial^{-m}.
\]
We shall write $W=W^{(1)}$ and $W_{ij}$ for $W^{(1)}_{ij}$ to simplify
notation,
then the simplest Sato equation is
\begin{equation}
\label{3.17a}
\frac{\partial P}{\partial x^{(k)}_{1}}=[\partial E_{kk},P]+[W,E_{kk}]P.
\end{equation}
In particular we have for
have for $i \neq k$:
\begin{equation}
\label{3.7.1}
\frac{\partial W_{ij}}{\partial x^{(k)}_{1}} = W_{ik} W_{kj} -
\delta_{jk} W^{(2)}_{ij}.
\end{equation}
The equation (\ref{3.17a}) is equivalent to the following equation for
$V=V^+(\alpha)$:
\begin{equation}
\label{3.17b}
\frac{\partial V}{\partial x^{(k)}_{1}}=(E_{kk}\partial +[W,E_{kk}])V.
\end{equation}

\section{Solutions of the Darboux-Egoroff system}

\noindent
Define
\begin{equation}
\label{4.1}
\gamma_{ij}(x)=W_{ij}^{(1)}(0,x)|_{x_k^{(i)}=c_k^{(i)}\ \text{for }k>1},
\end{equation}
where the $x_k^{(i)}$ for $k>0$ are chosen to be certain specific but at the
moment still unknown constants. From (\ref{3.7.1}) we already know
that
\begin{equation}
\label{4.2}
\frac{\partial \gamma_{ij}(x)}{\partial
x_1^{(k)}}=\gamma_{ik}(x)\gamma_{kj}(x)\quad i\ne k\ne j.
\end{equation}
This is the $n$-wave equation
if $i,j,k$ are distinct. The aim of this section is to construct
specific $\gamma_{ij}$'s which satisfy
\begin{equation}
\label{4.3}
\sum_{k=1}^n\frac{\partial \gamma_{ij}(x)}{\partial
x_1^{(k)}}=0,
\end{equation}
and
\begin{equation}
\label{4.4}
\gamma_{ij}(x)=\gamma_{ji}(x).
\end{equation}
In other words we want to find the rotation coefficients $\gamma_{ij}$
for the Darboux--Egoroff system (\ref{4.2})-(\ref{4.4}).
Sometimes we will assume an additional equation, viz.
\begin{equation}
\label{4.5}
\sum_{k=1}^nx_1^{(k)}\frac{\partial \gamma_{ij}(x)}{\partial
x_1^{(k)}}=-\gamma_{ij}(x),
\end{equation}
which means that $\gamma_{ij}$ has degree $-1$.
This equation holds for the so-called semisimple conformal invariant
Frobenius manifolds,
see \cite{Du1}.

The restriction
\begin{equation}
\label{4.6}
\sum_{k=1}^n\frac{\partial W_{ij}^{(1)}(0,x) }{\partial
x_1^{(k)}}=0,
\end{equation}
is a very natural restriction.
If we assume that
\begin{equation}
\label{4.7}
\sum_{k=1}^n\frac{\partial \tau(x) }{\partial
x_1^{(k)}}=0,
\end{equation}
then this clearly implies (\ref{4.6}).
Notice that one may even assume that
$\sum_{k=1}^n\frac{\partial \tau(x) }{\partial
x_1^{(k)}}=\lambda\tau(x)$,
but since we are in the polynomial case $\lambda$ must be $0$.
Equation (\ref{4.7}) means that $\tau$ (in the fermionic picture)
belongs to
the $GL_n(\mathbb{C}[t,t^{-1}])$-loop group orbit or even the
$SL_n(\mathbb{C}[t,t^{-1}])$-loop group orbit of $|0\rangle$ (see
\cite{KV} for more details). The homogeneous space for this group is
in fact the restricted Grassmannian
\[
\overline{Gr}=\{ W\in Gr_0|\sum_{k=1}^n tE_{kk}W\subset W\}.
\]
In fact $\tau$ satisfies (\ref{4.7}) if and only if
\begin{equation}
\label{4.8}
\sum_{k=1}^n
tE_{kk}W_\tau \subset W_\tau.
\end{equation}
Since  equation (\ref{4.7}) holds for $\tau$ we do not only find
equation
(\ref{4.6}) for $W^{(1)}(0,x)$, but we find that this equation holds
for all $W^{(s)}(\alpha,x)$'s and hence
\begin{equation}
\label{4.9a}
\sum_{k=1}^n\frac{\partial P^+(\alpha,x) }{\partial
x_1^{(k)}}=0.
\end{equation}
This means that we do not really have formal
pseudo-differential operators, but rather formal matrix-valued
Laurent series in $z^{-1}$. The Sato equation takes the following
simple form.
Let $P(z)=P^+(\alpha,x,z)$ then
\[
\frac{\partial P(z)}{\partial x^{(j)}_{k}} = -(P(z)E_{jj}
P(z)^{-1}z^k)_{-} P(z)
\]
and the simplest Sato'equation becomes
\[
\frac{\partial P(z)}{\partial x^{(k)}_{1}}=z[ E_{kk},P(z)]+[W,E_{kk}]P(z).
\]
Equation (\ref{3.17b}) turns into
\begin{equation}
\label{4.a}
\frac{\partial V(z)}{\partial x^{(k)}_{1}}=(zE_{kk} +[W,E_{kk}])V(z),
\end{equation}
where $V(z)=V^+(\alpha,x,z)$. Define $X=\sum_{j=1}^nx_1^{(j)}E_{jj}$, then
\begin{equation}
\label{4.aa}
\sum_{j=1}^n x^{(j)}_{1}\frac{\partial}{\partial x^{(j)}_{1}}V(z)
=(zX+[W,X])V(z).
\end{equation}
{}From now on we will only consider
tau-functions that are homogeneous with respect to the energy.
Notice that if $\text{energy }\tau=N$, then
$\text{energy }\tau_\alpha=N-\frac{1}{2}(\alpha|\alpha)$, in
particular $\text{energy }\tau_{\delta_i-\delta_j}
=\text{energy }\tau_0-1$.
Since the $\text{energy }x^{(j)}_k=k$, it is straightforward to check
that for $\alpha=0$,
\begin{equation}
\label{4.aaa}
\begin{aligned}
L_0&V(z)=z\frac{\partial V(z)}{\partial z},\quad\text{where}\\
L_0&=\sum_{j=1}^n\sum_{k=1}^\infty
kx_k^{(j)}\frac{\partial}{\partial x_k^{(j)}}.\\
\end{aligned}
\end{equation}

We will now describe a class of homogeneous tau-functions,
in the fermionic picture that satisfy (\ref{4.7}).
First choose two positive integers $m_1$ and $m_2$ such
that $m_1+m_2\le n$. Next choose $m_1$ positive integers $k_i$,
$1\le i\le m_1$ and $m_2$ positive integers $\ell_j$, $1\le j\le m_2$,
such that $\sum_{i=1}^{m_1} k_i -\sum_{j=1}^{m_2} \ell_j =0$.
Next choose $m_1$ linearly independent vectors
$a_i=(a_{i1},a_{i2},\ldots,a_{in})$ and $m_2$ linearly independent vectors
$b_j=(b_{j1},b_{j2},\ldots,b_{jn})$ in $\mathbb{C}^n$ such that
\begin{equation}
\label{4.9}
(a_i,b_j)=\sum_{k=1}^n a_{ik}b_{jk}=0\text{ for all }
1\le i\le m_1\text{ and } 1\le j\le m_2.
\end{equation}
Using lemma \ref{l1} we construct a $\tau\in\cal{O}_0$ as follows
\begin{equation}
\label{4.10}
\begin{aligned}
\tau=&
(\sum_p a_{1p} \psi^{+(p)}_{-k_1+\frac{1}{2}})
(\sum_p a_{1p} \psi^{+(p)}_{-k_1+\frac{3}{2}})
\cdots
(\sum_p a_{1p} \psi^{+(p)}_{-\frac{1}{2}})
(\sum_p a_{2p} \psi^{+(p)}_{-k_2+\frac{1}{2}})
(\sum_p a_{2p} \psi^{+(p)}_{-k_2+\frac{3}{2}})
\cdots\\
&
(\sum_p a_{2p} \psi^{+(p)}_{-\frac{1}{2}})
(\sum_p a_{3p} \psi^{+(p)}_{-k_3+\frac{1}{2}})
\cdots
(\sum_p a_{m_1,p} \psi^{+(p)}_{-\frac{1}{2}})
(\sum_p b_{1p} \psi^{-(p)}_{-\ell_1+\frac{1}{2}})
(\sum_p b_{1p} \psi^{-(p)}_{-\ell_1+\frac{3}{2}})
\cdots\\
&
(\sum_p b_{1p} \psi^{-(p)}_{-\frac{1}{2}})
(\sum_p b_{2p} \psi^{-(p)}_{-\ell_2+\frac{1}{2}})
\cdots
(\sum_p b_{m_2,p} \psi^{-(p)}_{-\frac{1}{2}})
|0\rangle .\\
\end{aligned}
\end{equation}
The point of the Grassmannian $W_\tau$ corresponding to this $\tau$
satisfies (\ref{4.8}).

The symmetry conditions (\ref{4.4}) of the $\gamma_{ij}$'s are not so natural.
Using (\ref{3.3.7}), it is equivalent to
\begin{equation}
\label{4.11}
\tau_{\delta_i-\delta_j}(x_1^{(\ell)},c_2^{(\ell)},c_3^{(\ell)},\ldots)=-
\tau_{\delta_j-\delta_i}(x_1^{(\ell)},c_2^{(\ell)},c_3^{(\ell)},\ldots).
\end{equation}
To achieve this result, we define an automorfism $\omega$ on $F$ as
follows:
\begin{equation}
\label{4.12}
\begin{aligned}
\omega(|0\rangle )&= |0\rangle\\
\omega(\psi_k^{\pm (i)}) &=c_i^{\pm 1}\psi_k^{\mp (i)},
\text{ with } 1\le i\le n \text{ and }c_i\in\mathbb{C}^\times.\\
\end{aligned}
\end{equation}
We will fix the $c_i$ later all to be equal to 1, but for the
moment we keep them arbitrary.
This gives
\begin{equation}
\label{4.13}
\omega(\alpha_k^{(i)})=-\alpha_k^{(i)}\text{ and }
\omega(Q_i^{\pm 1})=c_i^{\pm 1}Q_i^{\mp 1}.
\end{equation}
Using the boson-fermion correspondence this induces an automorfism on $B$,
which we will also denote by $\omega$,
\begin{equation}
\label{4.14}
\omega(x_k^{(i)})=-x_k^{(i)},\
\omega(\frac{\partial}{\partial x_k^{(i)}})=
-\frac{\partial}{\partial x_k^{(i)}},\
\omega(\delta_i)=-\delta_i\text{ and }
\omega(e^{\pm \delta_i})=c_i^{\pm 1}e^{\mp \delta_i}.
\end{equation}
Define for $\alpha=\sum_{j=1}^n p_i\delta_i\in M$,
$c_\alpha=\prod_{j=1}^n c_i^{p_i}$, then
\begin{equation}
\label{4.15}
\omega\left ( \sum_{\alpha\in M}\tau_\alpha (x)e^\alpha\right ) =
\sum_{\alpha\in M}c_\alpha\tau_\alpha (-x)e^{-\alpha}.
\end{equation}
We now want to find homogeneous tau-functions that satisfy
$\omega (\tau(x))=\lambda\tau(x)$ for some $\lambda\in\mathbb{C}^\times$.
Since $\omega^2(\tau_0(x))=\tau_0(x)$, $\lambda=1$ or $-1$.
{}From (\ref{4.15}) we deduce that
\begin{equation}
\label{4.16}
\tau_\alpha(x)=\lambda c_\alpha\tau_{-\alpha}(-x)
\end{equation}
and we want this for $\alpha\in\Delta$, of course after a specific choice of
constants  $x_k^{(i)}$'s for $k\ge 2$, to be equal to
$-\tau_\alpha(x)$.
Since we have assumed that
$\tau$ is homogeneous (in the energy), say that it has energy $N$,
then we can get rid of the $-x$ in
the right-hand side of (\ref{4.16}) if we put all $x_{2k}^{(i)}$'s
equal to zero. So define
\begin{equation}
\label{4.17}
\overline{\tau}(x)=\tau(x)|_{x_{2k}^{(i)}=0},
\end{equation}
then clearly (\ref{4.16}) turns into
\[
\overline{\tau}_\alpha(x) =
\lambda c_\alpha(-)^{N-\frac{1}{2}(\alpha|\alpha)}
\overline{\tau}_{-\alpha}(x).
\]
Because this also has to hold for $\alpha=0$, we obtain that
$\lambda=(-1)^N$ and hence $c_\alpha=1$ for all $\alpha\in\Delta$.
Thus $c_i=1$ for all $1\le i\le n$ or $c_i=-1$ for all $i$,
we may choose either of these two cases, for simplicity we choose
\[
c_i=1\text{ for all }1\le i\le n.
\]
With all these choices, we have finally that
\begin{equation}
\label{4.18}
\omega(\overline{\tau}_\alpha(x))=(-)^{N-\frac{1}{2}(\alpha|\alpha)}
\overline{\tau}_{-\alpha}(x).
\end{equation}

Return to the tau-functions of the form (\ref{4.10}). If such a $\tau$
satisfies (\ref{4.18}) and it
contains a factor $\sum_i a_{\ell i}\psi_k^{+(i)}$ for a certain $\ell$,
then it must also contain a factor  $\sum_j b_{m j}\psi_k^{-(j)}$.
Since
\[
\text{energy }
\left (\sum_i a_{\ell i}\psi_k^{+(i)}\right )
\left (\sum_j b_{m j}\psi_k^{-(j)}\right)
=-2k\in 2\mathbb{Z}+1,
\]
we must assume that  there exists an $m$ such that
\[
\begin{aligned}
\omega\left (
\left (\sum_i a_{\ell i}\psi_k^{+(i)}\right )
\left (\sum_j b_{m j}\psi_k^{-(j)}\right)
\right )
\ &=-
\left (\sum_j b_{m j}\psi_k^{+(j)}\right )
\left (\sum_i a_{\ell i}\psi_k^{-(i)}\right)\\
\ &=
-\left (\sum_i a_{\ell i}\psi_k^{+(i)}\right )
\left (\sum_j b_{m j}\psi_k^{-(j)}\right).
\end{aligned}
\]
So
\[
a_{\ell i}b_{mj}=a_{\ell j}b_{m i}\text{ for all }1\le i,j\le n
\]
and $b_m$ must be a multiple of $a_\ell$. Since the length of such a vector
does not matter much (only a scalar multiple of the whole tau-function),
we may assume that $a_\ell=b_m$ and
since also $(a_\ell,b_\ell)=0$ (see (\ref{4.9})),
we obtain that $a_\ell$ is an isotropic vector in $\mathbb{C}^n$.

Finally we conclude the following
\begin{proposition}
\label{p4.1}
Let $m$ be the integer part of $\frac{n}{2}$. Choose $m$ linearly
independent vectors
$a_i=(a_{i1},a_{i2},\ldots,a_{in})$  in $\mathbb{C}^n$ which span a maximal
isotropic subspace of $\mathbb{C}^n$, i.e.
\[
(a_i,a_j)=\sum_{k=1}^n a_{ik}a_{jk}=0\text{ for all }
1\le i,j\le m.
\]
Choose $m$ non-negative integers $k_i$, $1\le i\le m$
such that
\begin{equation}
\label{4.20b}
k_1\ge k_2\ge\ldots\ge k_m\ge 0,
\end{equation}
then
$\sigma(\tau)=\sum_{\alpha\in M}\tau_\alpha(x)e^\alpha$, with
\begin{equation}
\label{4.19}
\begin{aligned}
\tau=&
(\sum_p a_{1p} \psi^{+(p)}_{-k_1+\frac{1}{2}})
(\sum_p a_{1p} \psi^{+(p)}_{-k_1+\frac{3}{2}})
\cdots
(\sum_p a_{1p} \psi^{+(p)}_{-\frac{1}{2}})
(\sum_p a_{2p} \psi^{+(p)}_{-k_2+\frac{1}{2}})
(\sum_p a_{2p} \psi^{+(p)}_{-k_2+\frac{3}{2}})
\cdots\\
&(\sum_p a_{2p} \psi^{+(p)}_{-\frac{1}{2}})
(\sum_p a_{3p} \psi^{+(p)}_{-k_3+\frac{1}{2}})
\cdots
(\sum_p a_{mp} \psi^{+(p)}_{-\frac{1}{2}})
(\sum_p a_{1p} \psi^{-(p)}_{-k_1+\frac{1}{2}})
(\sum_p a_{1p} \psi^{-(p)}_{-k_1+\frac{3}{2}})
\cdots\\
&
(\sum_p a_{1p} \psi^{-(p)}_{-\frac{1}{2}})
(\sum_p a_{2p} \psi^{-(p)}_{-k_2+\frac{1}{2}})
\cdots
(\sum_p a_{mp} \psi^{-(p)}_{-\frac{1}{2}})
|0\rangle ,\\
\end{aligned}
\end{equation}
satisfies the $n$-component KP hierarchy (\ref{3.3.1}) and
\[
\omega(\tau)=(-)^{k_1+k_2+\cdots+k_m}\tau.
\]
 Moreover
\begin{equation}
\label{4.20a}
\text{\rm energy }
\tau_{\alpha}(x)=k^2_1+k^2_2+\cdots+k^2_m-\frac{1}{2}(\alpha |\alpha),
\end{equation}
\[
\sum_{j=1}^n \frac{\partial \tau_\alpha(x)}{\partial x_1^{(j)}}=0
\]
and
\[
\overline{\tau}_\alpha(x)=(-)^{\frac{1}{2}(\alpha|\alpha)}
\overline{\tau}_{-\alpha}(x),
\]
where $ \overline{\tau}$ is defined by (\ref{4.17}).
\end{proposition}
Notice that the restriction (\ref{4.20b}) is not essential, but we may
assume it without loss of generality.
Since the energy is nowhere  negative, formula
(\ref{4.20a}) gives a restriction for $\text{supp }\tau$.

It is not difficult to prove that the perfect wedge $\tau$, given by
(\ref{4.19}), is also a highest weight vector for the
$W_{1+\infty}$-algebra generated by
\[
J^{(\ell+1)}(z)=\sum_{k\in\mathbb{Z}}J_k^{(\ell+1)}z^{-k-\ell-1}=
\sum_{j=1}^n :\psi^{+(j)}(z)\frac{\partial^\ell
\psi^{-(j)}(z)}{\partial z^\ell}:\quad \ell=0,1,2,\ldots,
\]
i.e.,
\[
J_k^{(\ell+1)}\tau=\delta_{k0}c_{\ell}(k_1,k_2,\ldots,k_m)\tau\quad\text{for
}k\ge 0.
\]
Here $c_\ell\in\mathbb{C}$ only depend on the integers $k_1,k_2,\ldots,k_m$.
This induces the following restriction on
$W_\tau\in Gr_0$:
\[
\sum_{j=1}^n t^{k+\ell}\left (\frac{\partial}{\partial t}\right )^\ell
E_{jj}W_\tau\subset W_\tau\quad\text{for all }k,\ell=0,1,2,\ldots
\]

If we now rewrite the element (\ref{4.19}) as a perfect wedge,
we can use proposition \ref{p1} to determine $\tau_\alpha$ for
$\alpha=0$ or $\alpha\in\Delta$.
Add to the vectors
$a_i$, $1\le i\le m$ vectors $a_j$, $m+1\le j\le n$ such that
they form a basis of $\mathbb{C}^n$, which satisfies
\begin{equation}
\label{4.20}
(a_\ell,a_k)=\delta_{k+\ell,2m+1}+\delta_{k+\ell,4m+2}
\text{ for all }1\le k,\ell \le n.
\end{equation}
Define
\begin{equation}
\label{4.20d}
k_{2m+1-i}=-k_i\text{ for } 1\le i\le m.
\end{equation}
Then
the $\tau$ given by (\ref{4.19}) is upto a scalar multiple equal to the
following
perfect wedge
\[
A_{-\frac{1}{2}}\wedge A_{-\frac{3}{2}}\wedge A_{-\frac{5}{2}}\wedge\cdots,
\]
with
\begin{equation}
\label{4.21}
\begin{aligned}
A_{-qk_1-(k_1+k_2+\cdots+k_{q-1})-\ell}&=
\sum_{j=1}^n a_{qj}v_\ell^{(j)}
\text{ with }1\le q\le 2m-1 \text{ and }
-k_1+\frac{1}{2}\le \ell \le k_q-\frac{1}{2},\\
A_{-(2m)-(k_1+k_2+\cdots k_{2m-1})-\ell}&=
\sum_{j=1}^n a_{n,j}v_\ell^{(j)}\text{ with }
-k_1+\frac{1}{2}\le \ell\le -\frac{1}{2}
\text{ this only if }n=2m+1,\\
A_q&=v_q\text{ for }q< -nk_1-k_2-\cdots-k_{2m-1}.\\
\end{aligned}
\end{equation}
Now using (\ref{2.10}), this is equal to
\begin{equation}
\label{4.22}
\begin{aligned}
A_{-qk_1-(k_1+k_2+\cdots+k_{q-1})-\ell}&=
\sum_{j=1}^n a_{qj}v_{n\ell-\frac{1}{2}(n-2j+1)}
\text{ with }1\le q\le 2m-1 \text{ and }
-k_1+\frac{1}{2}\le \ell \le k_q-\frac{1}{2},\\
A_{-(2m)-(k_1+k_2+\cdots k_{2m-1})-\ell}&=
\sum_{j=1}^n a_{n,j}v_{n\ell-\frac{1}{2}(n-2j+1)}\text{ with }
-k_1+\frac{1}{2}\le \ell\le -\frac{1}{2}
\text{ this only if }n=2m+1\\
A_q&=v_q\text{ for }q< -nk_1-k_2-\cdots-k_{2m-1}.\\
\end{aligned}
\end{equation}
Using proposition \ref{p1}, one easily deduces that the to (\ref{4.19})
corresponding $\tau_0$
is given by
\[
\begin{aligned}
\tau_0=
\det (
&
\sum_{q=1}^{2m-1}
\sum_{j=1}^n \sum_{i=1}^{k_1}
 \sum_{\ell=-i}^{ k_q-1}
a_{qj}S_{\ell+i}(x^{(j)})
E_{j-in-\frac{1}{2},
-qk_1-(k_1+k_2+\cdots+k_{q-1})-\ell-\frac{1}{2}}\\
\ &
+\delta_{(-1)^n,-1}
\sum_{j=1}^n
\sum_{i=1}^{k_1}
\sum_{\ell=-i}^{-1}
a_{n,j} S_{\ell+i}(x^{(j)})
E_{j-in-\frac{1}{2},
-(2m)-(k_1+k_2+\cdots k_{2m-1})-\ell-\frac{1}{2}})\\
\end{aligned}
\]
and $\tau_{\delta_r-\delta_s}$ for $1\le r,s\le n$ is equal to
the determinant of $\tau_0$, but then with the $(s-n-\frac{1}{2})$-th row
replaced by
\[
\sum_{q=1}^{2m-1}
 \sum_{\ell=0}^{ k_q-1}
a_{qr}S_{\ell}(x^{(r)})
E_{s-n-\frac{1}{2},
-qk_1-(k_1+k_2+\cdots+k_{q-1})-\ell-\frac{1}{2}}
\]
Now change the indices and we obtain
\begin{theorem}
\label{t4.1}
Let $\tau$ be given by (\ref{4.19}), and let
$\sigma(\tau)=\sum_{\alpha\in M}\tau_\alpha(x)e^\alpha$, then upto a
common scalar factor
\begin{equation}
\label{4.23}
\begin{aligned}
\tau_0=
\det (
&
\sum_{q=1}^{2m-1}
\sum_{j=1}^n \sum_{i=1}^{k_1}
 \sum_{\ell=1-k_q}^{i}
a_{qj}S_{i-\ell}(x^{(j)})
E_{in-j +1,
qk_1+(k_1+k_2+\cdots+k_{q-1})-\ell+1}\\
\ &
+\delta_{(-1)^n,-1}
\sum_{j=1}^n
\sum_{i=1}^{k_1}
\sum_{\ell=1}^{i}
a_{n,j} S_{i-\ell}(x^{(j)})
E_{in-j+1,
(2m)+(k_1+k_2+\cdots k_{2m-1})-\ell+1})\\
\end{aligned}
\end{equation}
and $\tau_{\delta_r-\delta_s}$ for $1\le r,s\le n$ is equal to
the determinant of $\tau_0$, but then with the $(n+1-s)$-th row
replaced by\begin{equation}
\label{4.24}
\sum_{q=1}^{2m-1}
 \sum_{\ell=0}^{ k_q-1}
a_{qr}S_{\ell}(x^{(r)})
E_{n+1-s,
qk_1+(k_1+k_2+\cdots+k_{q-1})+\ell+1},
\end{equation}
where the $a_\ell$, $1\le \ell\le n$, satisfy (\ref{4.20}) and the
$k_j$, $m+1\le j\le 2m$ are given by (\ref{4.20d})
Moreover, the
\begin{equation}
\label{4.25}
\overline{\gamma}_{rs}(x)=\begin{cases}
\epsilon_{sr}\frac{\overline{\tau}_{\delta_r-\delta_s}(x)}
{\overline{\tau}_0(x)},
&\text{if }1\le r,s\le n\text{ and }r\ne s,\\
-\frac{\partial \log \overline{\tau}_0(x)}{\partial x_1^{(r)}},
&\text{if }1\le r,s\le n\text{ and }r= s,\\
\end{cases}
\end{equation}
satisfy the Darboux-Egoroff system (\ref{4.2})-(\ref{4.4}). If we
define
\begin{equation}
\label{4.26}
\gamma_{rs}(x)=\overline{\gamma}_{rs}(x)|_{x_k^{(i)}=0\text{ for all }k>1},
\end{equation}
then these elements satisfy (\ref{4.2})-(\ref{4.5}).
\end{theorem}

Let $f(t)=\sum_i f_i(t)e_i$ and $ g(t)=\sum_i g_i(t)e_i$ be two
elements in  $H$, define the following bilinear form
\begin{equation}
\label{4.30}
B(f,g)=\text{\rm Res}_{t=0}\sum_{i=1}^n f_i(t)g_i(t).
\end{equation}
Then the orthogonal restricted Grassmannian is
\begin{equation}
\label{4.31}
\widehat{Gr}=\{ W\in \overline{Gr}|B(W,W)=0\}.
\end{equation}
All $W\in\widehat{Gr}$ are maximal isotropic subspaces with respect to
$B(\cdot,\cdot)$.
This Grassmannian is the homogeneous space for the
$O_n(\mathbb{C}[t,t^{-1}])$-loop group. The
$O_n(\mathbb{C}[t,t^{-1}])$-orbit of $|0\rangle$ corresponds exactly
to this Grassmannian (see e.g. \cite{PS}).
Notice that all the $W_\tau$'s corresponding to the tau-functions
given by (\ref{4.19}) exactly satisfy this condition. Hence the
tau-functions we have constructed to solve the Darboux-Egoroff system
are in fact homogeneous tau-functions in the
$O_n(\mathbb{C}[t,t^{-1}])$-orbit of $|0\rangle$.
If we consider the affine Lie algebra
$gl_{n}({\mathbb C})^{\wedge}$ with central charge $1$, defined by
(\ref{2.1.7}),
then the special orthogonal Lie algebra $so_n(\mathbb{C})^\wedge$ is
given by
\[
so_n(\mathbb{C})^\wedge=\{ x\in gl_{n}({\mathbb C})^{\wedge}|\omega (x)=x\}.
\]
Recall that $\omega(\psi^{\pm(i)}_k)=\psi^{\mp(i)}_k$.
The Grassmannian
$\widehat{Gr}$ has two connected components, which are distinguished by
the parity of the dimension of the kernel of the projection $W\to
H_0$. Depending on the energy of our (homogeneous) tau-function,
$\omega(\tau)=(-)^{\text{energy }\tau}\tau$, the space $W_\tau$
belongs to one of these two components.

It is obvious, from the above description and from the construction of
the tau-functions given by (\ref{4.19}), that the orthogonal group
$O_n$ acts on these tau-functions and hence on the rotation
coefficients. One has

\begin{proposition}
\label{p4.2}
The orthogonal group $O_n$ acts on the rotation coefficients of
Theorem \ref{t4.1}. Let $X=(X_{ij})_{1\le i,j\le n}\in O_n$, then
replacing $a_{ij}$, $1\le i,j\le n$, (even if $a_{ij}=0$) by
$\sum_{\ell=1}^nX_{j\ell}a_{i\ell}$ in (\ref{4.23}) and (\ref{4.24})
gives a new solution of the Darboux-Egoroff system.
\end{proposition}

\section{Semisimple Frobenius manifolds}

Let $\gamma_{ij}(x)$, $1\le i,j\le n$,
be a solution of the Darboux-Egoroff system.
If we can find $n$ linearly independent vector functions
$\psi_j=\psi_j(x)=^t\!\!(\psi_{1j},\psi_{2j},\ldots,\psi_{nj})$ such that
\begin{equation}
\label{5.1}
\begin{aligned}
\frac{\partial\psi_{ij}}{\partial x_1^{(k)}}&=
\gamma_{ik}\psi_{kj},\qquad k\ne i,\\
\sum_{k=1}^n
\frac{\partial\psi_{ij}}{\partial x_1^{(k)}}&= 0,\\
\end{aligned}
\end{equation}
then they determine under certain conditions (locally)
a semisimple (i.e. massive) Frobenius manifold (see \cite{Du1}, \cite{Du2}).

Recall from (\ref{4.a}), that the wave function
$V(z)=V^+(0,x,z)$ corresponding to the tau-functions
of proposition \ref{p1} and theorem \ref{t4.1} satisfy
\begin{equation}
\label{5.2}
\begin{aligned}
\frac{\partial V_{ij}(z)}{\partial x_1^{(k)}}&=
W_{ik}V_{kj}(z),\qquad k\ne i,\\
\sum_{k=1}^n
\frac{\partial V_{ij}(z)}{\partial x_1^{(k)}}&= zV_{ij}(z).\\
\end{aligned}
\end{equation}
Comparing (\ref{5.1}) and (\ref{5.2}), one would like to take
$z=0$ in (\ref{5.2}), however this does not make sense.
There is a way to use the wave function $V(z)$ to construct
the $\psi_{ij}$'s of (\ref{5.1}).
Suppose that we have a tau-function of the form (\ref{4.19}),
with the corresponding $k_q$'s, $1\le q\le n$,
(in the case that $n$ is odd, we define $k_n=0$) and $a_{qj}$'s
$1\le q,j\le n$. Let
\begin{equation}
\label{5.3}
X_q(t)=\sum_{j=1}^n a_{qj}t^{-k_q-1}e_j\in H,\qquad 1\le q\le n,
\end{equation}
then it easy to check that
\[
W_\tau +\mathbb{C}X_q(t)\ne W_\tau
\text{  and  }
W_\tau +\mathbb{C}tX_q(t)=W_\tau.
\]
Hence,
\begin{equation}
\label{5.4}
\sum_{j=1}^n a_{qj}\psi^{+(j)}_{-k_q-\frac{1}{2}}\tau\ne 0\quad
\text{ and }\quad
\sum_{j=1}^n a_{qj}\psi^{+(j)}_{-k_q+\frac{1}{2}}\tau= 0.
\end{equation}
We rewrite this as follows
\begin{equation}
\label{5.5a}
\text{Res}_{z=0}\sum_{j=1}^n a_{qj}z^{-k_q-1}\psi^{+(j)}(z)\tau\ne 0\quad
\text{  and  }\quad
\text{Res}_{z=0}\sum_{j=1}^n a_{qj}z^{-k_q}\psi^{+(j)}(z)\tau= 0.
\end{equation}
{}From which we deduce that
\[
\begin{aligned}
\text{Res}_{z=0}\sum_{j=1}^n a_{qj}z^{-k_q-1}
z^{1-\delta_{ij}}
e^{\sum_{\ell=1}^\infty x_\ell^{(j)}z^\ell}
e^{-\sum_{\ell=1}^\infty \frac{\partial}{\partial x^{(j)}_\ell }
\frac{z^{-\ell}}{\ell}}\tau_{\delta_i-\delta_j}(x)&\ne 0\quad\text{  and}\\
\text{Res}_{z=0}\sum_{j=1}^n a_{qj}z^{-k_q}
z^{1-\delta_{ij}}
e^{\sum_{\ell=1}^\infty x_\ell^{(j)}z^\ell}
e^{-\sum_{\ell=1}^\infty \frac{\partial}{\partial x^{(j)}_\ell }
\frac{z^{-\ell}}{\ell}}\tau_{\delta_i-\delta_j}(x)&= 0.\\
\end{aligned}
\]
Dividing this by $\tau_0(x)$ we obtain
\begin{equation}
\label{5.5}
\text{Res}_{z=0}\sum_{j=1}^n a_{qj}z^{-k_q-1}  V_{ij}(z)\ne 0\quad\text{
and  }\quad
\text{Res}_{z=0}\sum_{j=1}^n a_{qj}z^{-k_q} V_{ij}(z)=0.
\end{equation}
Now define for $1\le i,q\le n$
\begin{equation}
\label{5.6}
\Psi_{iq}= \text{Res}_{z=0}\sum_{j=1}^n a_{qj}z^{-k_q-1}  V_{ij}(z),
\end{equation}
then it is straighforward to check, using (\ref{5.2}) and (\ref{5.5}) that
\begin{equation}
\label{5.7}
\begin{aligned}
\frac{\partial\Psi_{ij}}{\partial x_1^{(k)}}&=
W_{ik}\Psi_{kj},\qquad k\ne i,\\
\sum_{k=1}^n
\frac{\partial\Psi_{ij}}{\partial x_1^{(k)}}&= 0.\\
\end{aligned}
\end{equation}
Notice that the vector functions
$\Psi_q=^t\! (\Psi_{1q},\Psi_{2q},\ldots,\Psi_{nq})$ are ``eigenfunctions'' of
the KP
hierachy which lie in the kernel of $L$.
{}From all this we finally obtain the following
\begin{theorem}
\label{t5.1}
Let $V(z)=V^+(0,x,z)$ be the wave function corresponding to the
tau-function of (\ref{4.19}) with $a_{qj}$, $1\le q,j\le n$
and $k_\ell$, $1\le \ell\le 2m$, as given in theorem \ref{t4.1}
and $k_n=0$ if $n$ is odd. Denote by
\begin{equation}
\label{5.8a}
\begin{aligned}
\psi_{iq}&= \text{\rm Res}_{z=0}\lambda_q \sum_{j=1}^n a_{qj}z^{-k_q-1}
V_{ij}^+(0,x,z)
|_{x_{k}^{(\ell)}=0 \text{ for all }k>1},\\
\overline{\psi}_{iq}&= \text{\rm Res}_{z=0}\lambda_q \sum_{j=1}^n
a_{qj}z^{-k_q-1}  V_{ij}^+(0,x,z)
|_{x_{2k}^{(\ell)}=0 \text{ for all }k},\\
\end{aligned}
\end{equation}
where $1\le q\le n$ and
$\lambda_q\in\mathbb{C}^\times$.
Then these $\psi_{iq}$'s satisfy the equations (\ref{5.1}), with
$\gamma_{ij}$ given by (\ref{4.25}) and the
formula's
\begin{equation}
\label{5.8}
\begin{aligned}
\eta_{ii}&=\psi_{i1}^2,\\
\eta_{\alpha\beta}&=\sum_{i=1}^n \psi_{i\alpha}\psi_{i\beta},\\
\frac{\partial t_\alpha}{\partial x_1^{(i)}}&=\psi_{i1}\psi_{i\alpha},\\
c_{\alpha\beta\gamma}&=\sum_{i=1}^n
\frac{\psi_{i\alpha}\psi_{i\beta}\psi_{i\gamma}}{\psi_{i1}},\\
\end{aligned}
\end{equation}
with $t_\alpha=\sum_{\epsilon=1}^n\eta_{\alpha\epsilon}t^\epsilon$,
determine (locally) a massive Frobenius manifold on the
domain $x_1^{(i)}\ne x_1^{(j)}$ and
$\psi_{11}\psi_{21}\cdots\psi_{n1}\ne 0$.
The $\overline{\psi}_{iq}$'s also satisfy (\ref{5.1}), but now with the
$\gamma_{ij}$
replaced by $\overline\gamma_{ij}$ of (\ref{4.26}). The equations
(\ref{5.8}) for these $\overline{\psi}_{ij}$'s also determine
a semisimple Frobenius manifold.
\end{theorem}
{\bf Proof}. Formula (\ref{5.8}) is a direct consequence of
the following proposition , see \cite{Du3} (cf. \cite{Du1} and
\cite{Du2}) for more details.
\hfill$\square$\break
\begin{proposition}
\label{p5.1a}
Let $X=\sum_{i=1}^n x_1^{(i)}E_{ii}$,
$\Gamma=(\gamma_{ij})_{1\le i,j\le n}$,
${\cal{V}}=[\Gamma,X]$ and ${\cal{V}}_k=[\Gamma,E_{kk}]$, then
${\cal{V}}=({\cal{V}}_{ij})_{1\le i,j\le n}$ is anti-symmetric
and satisfies
\begin{equation}
\label{V}
\frac{\partial{\cal{V}}}{\partial x_1^{(k)}}=
[{\cal{V}}_k,{\cal{V}}]
\end{equation}
and also
\begin{equation}
\label{5.9}
\begin{aligned}
{\cal V} \psi_q&=\sum_{j=1}^n x_1^{(j)}\frac{\partial \psi_q}{\partial
x_1^{(j)}}
=k_q\psi_q,\\
\frac{\partial\psi_q}{\partial x_1^{(k)}}&={\cal{V}}_k\psi_q,\\
\end{aligned}
\end{equation}
for $\psi_q=^t\!\! (\psi_{1q},\psi_{2q},\ldots,\psi_{nq})$
\end{proposition}
{\bf Proof}.
The equation (\ref{V}) follows from
(\ref{4.2}), (\ref{4.3}) and the fact that $\Gamma$ is symmetric.
We prove (\ref{5.9})  as follows. Let ${\cal V}$ act
$\psi_q$, Using (\ref{4.aa}) and (\ref{5.6}) one deduces
\[
{\cal V }\psi_q=
\sum_{j=1}^n x_1^{(j)}\frac{\partial \psi_q}{\partial
x_1^{(j)}}.
\]
Since $\psi_q$ is independent of $x^{(j)}_k$ for all $k>1$, we can use
(\ref{4.aaa}),
to rewrite this as follows
\[
\begin{aligned}
\sum_{j=1}^n x_1^{(j)}\frac{\partial \psi_{iq}}{\partial
x_1^{(j)}}=&\text{\rm Res}_{z=0}\lambda_q \sum_{j=1}^n a_{qj}z^{-k_q-1}
z\frac{\partial }{\partial z}\left(
V_{ij}^+(0,x,z)|_{x_{k}^{(\ell)}=0 \text{ for all }k>1}\right )
\\
=&\text{\rm Res}_{z=0}\lambda_q \sum_{j=1}^n a_{qj}\left (
\frac{\partial }{\partial z}z^{-k_q} +k_qz^{-k_q-1}\right )\left(
V_{ij}^+(0,x,z)|_{x_{k}^{(\ell)}=0 \text{ for all }k>1}\right )
\\
=&k_q\text{\rm Res}_{z=0}\lambda_q \sum_{j=1}^n a_{qj}
z^{-k_q-1}
V_{ij}^+(0,x,z)|_{x_{k}^{(\ell)}=0 \text{ for all }k>1}
\\
=&k_q\psi_{iq}.
\end{aligned}
\]
The second equation of (\ref{5.9}) can be proved in a similar way,
using (\ref{4.a}). \hfill$\square$\break
{}From (\ref{5.9}) we determine the degrees $d_1,d_2,\ldots,d_n$ and $d$
(resp. $d_F$) of the corresponding $t^\alpha$
\begin{equation}
\label{Ds}
d_1=1,\ d_\alpha=1+k_1-k_\alpha,\ 2\le \alpha\le n,\ d=-2k_1\
\text{and }d_F=3+2k_1.
\end{equation}
With our choice of $k_\alpha$ we have
\[
d_\alpha+d_{2m+1-\alpha}=2-d,\ 1\le \alpha\le m\
\text{and }d_n=1+k_1\ \text{if }n=2m+1\ \text{is odd}.
\]

Notice that if we
define
\begin{equation}
\label{A1}
\Phi(z)=V(0,x,z)|_{x_k^{(\ell)}=0\ \text{for all }k>1},
\end{equation}
then $\Phi(z)$ satisfies
\begin{equation}
\label{A2}
\begin{aligned}
z\frac{\partial \Phi(z)}{\partial z}&=\sum_{j=1}^n x_1^{(j)}
\frac{\partial \Phi(z)}{\partial
x_1^{(j)}}=
(zX+{\cal{V}})\Phi(z),\\
\frac{\partial \Phi(z)}{\partial x_1^{(k)}}&=
(zE_{kk}+{\cal{V}}_k)\Phi(z).\\
\end{aligned}
\end{equation}
\begin{theorem}
\label{t5.2a}
Let $\Psi=(\psi_{ij})_{1\le i,j\le n}$ and define
$\xi(z)=\ \!\!\!^t\!\Psi\Phi(z)=\eta\Psi^{-1}\Phi(z)$,
${\cal {U}}=\eta\Psi^{-1}X\Psi\eta^{-1}$,
$\mu=-\eta\Psi^{-1}{\cal{V}}\Psi\eta^{-1}=\sum_{i=1}^n k_iE_{ii}$
and $\Pi_i=\eta\Psi^{-1}E_{ii}\Psi\eta^{-1}$, then
$\eta(^t\cal {U})={\cal {U}}\eta$, $\mu\eta+\eta\mu=0$
and
\begin{equation}
\label{xi}
\begin{aligned}
z\frac{\partial\xi(z)}{\partial z}&=(z{{\cal{U}}}-\mu)\xi(z),\\
\sum_{j=1}^n x_1^{(j)}
\frac{\partial \xi(z)}{\partial
x_1^{(j)}}&=z{\cal{U}}\xi(z),\\
\frac{\partial\xi(z)}{\partial x_1^{(k)}}&=z\Pi_k\xi(z),\\
\frac{\partial \xi(z)}{\partial
t^\alpha}&=zC_\alpha\xi(z),\\
\end{aligned}
\end{equation}
where $C_\alpha=\sum_{\beta,\gamma=1}^n c_{\alpha\beta}^\gamma
E_{\beta\gamma}$.
\end{theorem}
{\bf Proof.} All formula's except the last one of (\ref{xi}) follows
immediately
from
(\ref{V}), (\ref{5.9}),
(\ref{A2}) and the fact that $\ ^t\!\Psi\Psi=\eta$. Use the last
formula of (\ref{5.8}), $c_{\alpha\beta}^\gamma=\sum_{\epsilon=1}^n
c_{\beta\alpha\epsilon}\eta^{\epsilon\gamma}$ and
$\frac{\partial x_1^{(i)}}{\partial t^\alpha}
=\frac{\psi_{i\alpha}}{\psi_{i1}}$ to rewrite
\[
\begin{aligned}
\frac{\partial \xi}{\partial
t^\alpha}=&\sum_{i=1}^n\frac{\partial x_1^{(i)}}{\partial
t^\alpha}
\frac{\partial \xi}{\partial
x_1^{(i)}}\\
=&z\sum_{i=1}^n\frac{\partial x_1^{(i)}}{\partial
t^\alpha}\eta\Psi^{-1}E_{ii}\Psi\eta^{-1}\xi\\
=&z^t\!\Psi\sum_{i=1}^n\frac{\psi_{i\alpha}}{\psi_{i1}}E_{ii}\Psi\eta^{-1}\xi\\
=&zC_\alpha\xi.\\
\end{aligned}
\]
This finishes the proof of the theorem.\hfill$\square$\break

As in \cite{Du2},\cite{Du3} we can reformulate (\ref{V}) as an
$\{x_1^{(i)}\}_{1\le i\le n}$-dependent
commuting Hamiltonian system
\[
\frac{\partial{\cal{V}}}{\partial x_1^{(k)}}=
\{ {\cal{V}}, H_k({\cal{V}},X)\},
\]
with quadratic Hamiltonians
\begin{equation}
\label{5.12}
H_i({\cal{V}},X)=\frac{1}{2}\sum_{j\ne i}
\frac{{\cal{V}}_{ij}{\cal{V}}_{ji}}{x_1^{(i)}-x_1^{(j)}}=
\frac{1}{2}\sum_{j\ne i}
\gamma_{ij}\gamma_{ji}(x_1^{(i)}-x_1^{(j)})
\end{equation}
with respect to the standard Poisson bracket on $so_n$:
\[
\{{\cal{V}}_{ij},{\cal{V}}_{k\ell}\}=
\delta_{jk}{\cal{V}}_{i\ell}-\delta_{ik}{\cal{V}}_{j\ell}
+\delta_{i\ell}{\cal{V}}_{jk}-\delta_{j\ell}{\cal{V}}_{ik}.
\]
Now consider the $1$-form
\begin{equation}
\label{5.13}
\sum_{i=1}^n H_i({\cal{V}},X)dx_1^{(i)}.
\end{equation}
Since it is closed for any such $\cal V$(see \cite{Du1}, \cite{Du2}), there
exists a function
$\tau_I(X)$, the isomonodromy tau-function, such that
\begin{equation}
\label{5.14}
d\log \tau_I(X)=\sum_{i=1}^n H_i({\cal{V}},X)dx_1^{(i)}.
\end{equation}
Using (\ref{4.2}), we rewrite $H_i({\cal{V}},X)$ as follows. Let
$\tilde\tau_0(X)=\tau_0(x)|_{x_k^{(\ell)}=0\ \text{for all }k>1}$,
then
\[
\begin{aligned}
H_i({\cal{V}},X)=&
\frac{1}{2}\sum_{j\ne i}
\gamma_{ij}\gamma_{ji}(x_1^{(i)}-x_1^{(j)})\\
=&\frac{1}{2}\sum_{j\ne i}
\frac{\partial\gamma_{ii}}{\partial x_1^{(j)}}(x_1^{(i)}-x_1^{(j)})\\
=&\frac{1}{2}\sum_{j=1}^n
x_1^{(i)}\frac{\partial\gamma_{ii}}{\partial x_1^{(j)}}
-\frac{1}{2}\sum_{j=1}^n
x_1^{(j)}\frac{\partial\gamma_{ii}}{\partial x_1^{(j)}}
\\
=&-\frac{1}{2}\sum_{j=1}^n
x_1^{(j)}\frac{\partial\gamma_{ii}}{\partial x_1^{(j)}}
\\
=&
\frac{1}{2}\sum_{j=1}^n x_1^{(j)}\frac{\partial}{\partial x_1^{(j)}}
\frac{\partial}{\partial x_1^{(i)}} \left (\log \tilde\tau_0(X)\right )\\
=&\frac{1}{2}\frac{\partial}{\partial x_1^{(i)}}
\left(
\sum_{j=1}^n
x_1^{(j)}\frac{\partial}{\partial x_1^{(j)}}
\left (\log \tilde\tau_0(X)\right )\right )
-\frac{1}{2} \frac{\partial}{\partial x_1^{(i)}}
\left (\log \tilde\tau_0(X)\right )\\
=&-\frac{1}{2} \frac{\partial}{\partial x_1^{(i)}}
\left (\log \tilde\tau_0(X)\right ).\\
\end{aligned}
\]
Hence
\begin{equation}
\label{5.15}
d\log \tau_I(X)=-\frac{1}{2}d\log\tilde\tau_0(X).
\end{equation}
Dubrovin and Zhang defined in \cite{DuZ} a Gromov-Witten type $G$-function
of a Frobenius manifold as follows
\begin{equation}
\label{5.16}
\begin{aligned}
G&=\log\left (\frac{\tau_I}{J^{\frac{1}{24}}}\right ),
\quad\text{where}\\
J&=\det\left (\frac{\partial t^\alpha}{\partial x_1^{(i)}}\right )=
\log\left( \psi_{11}\psi_{21}\cdots\psi_{n1}\right ).\\
\end{aligned}
\end{equation}
We can explicitly determine this function in the cases of the Frobenius
manifolds corresponding to theorem \ref{t5.1}.
\begin{theorem}
\label{t5.2}
Let $\tau$ be given by (\ref{4.19}) and let $\psi_{i1}$ be defined as in
(\ref{5.8a}).
Let $\tilde\tau_0(X)=\tau_0(x)|_{x_k^{(\ell )}=0\ \text{for all
}k>1}$, i.e.,
\begin{equation}
\begin{aligned}
\tilde\tau_0(X)=
\det (
&
\sum_{q=1}^{2m-1}
\sum_{j=1}^n \sum_{i=1}^{k_1}
 \sum_{\ell=1-k_q}^{i}
a_{qj}\frac{(x_1^{(j)})^{i-\ell}}{(i-\ell)!}
E_{in-j +1,
qk_1+(k_1+k_2+\cdots+k_{q-1})-\ell+1}\\
\ &
+\delta_{(-1)^n,-1}
\sum_{j=1}^n
\sum_{i=1}^{k_1}
\sum_{\ell=1}^{i}
a_{n,j} \frac{(x_1^{(j)})^{i-\ell}}{(i-\ell)!}
E_{in-j+1,
(2m)+(k_1+k_2+\cdots k_{2m-1})-\ell+1})\\
\end{aligned}
\end{equation}
Then upto an additive scalar factor
\begin{equation}
\label{5.17}
G=-\frac{1}{2}\log\tilde\tau_0(X)-\frac{1}{24}
\log\left( \psi_{11}\psi_{21}\cdots\psi_{n1}\right ).
\end{equation}
Moreover,
\[
\sum_{j=1}^nx_1^{(j)}\frac{\partial G}{\partial x_1^{(j)}}=\gamma G,
\]
where
\begin{equation}
\label{gamma}
\gamma=-\frac{1}{4}\sum_{j=1}^nk_j^2-\frac{nk_1}{24}
\end{equation}
and
\[
\frac{\partial}{\partial x_1^{(i)}} \frac{\partial}{\partial x_1^{(j)}}
\left (\log \tilde\tau_0(X)\right )=-\gamma_{ij}^2\qquad i\ne j,
\]
where $\gamma_{ij}$ is defined by formula (\ref{4.26}).
\end{theorem}

\section{An example}
In this section we describe the simplest example in more detail.
Let $n=2m$, respectively $n=2m+1$ if n is even respectively odd.
Since the choices of the order of $k_1,k_2\ldots k_m\in\mathbb{Z}$ is rather
arbitrary, we choose for simplicity of notation and calculation
$k_1=-k_{n}=-1$ and all other $k_i=0$.
Hence $d_1=1,\ d_n=-1,\ d_\alpha=0,\ \alpha\ne 1,n,\ d=2$ and $d_F=1$
Choose vectors
$a_i=(a_{i1},a_{i2},\ldots,a_{in})$, such that
\[
(a_i,a_j)=\delta_{i+j,n+1}.
\]
Then
\[
\tau_0=\sum_{j=1}^n a_{ni}^2u_i\
\text{and}\ \ \tau_{\delta_i-\delta_j}=-\tau_{\delta_j-\delta_i}=a_{ni}a_{nj}\
\text{for }
i< j,
\]
where we use the notation $u_i=x_1^{(i)}$.
Hence,
\[
\gamma_{ij}=-\frac{a_{ni}a_{nj}}{\sum_{j=1}^n a_{ni}^2u_i}\ \text{for
}1\le i,j\le n
\]
and the wave function is equal to
\[
V(z)=\left (I-\frac{1}{\tau_0}\sum_{i,j=1}^n
a_{ni}a_{nj}E_{ij}z^{-1}\right )
\sum_{\ell=1}^n
\sum_{k=0}^\infty
S_k(x^{(\ell)})E_{\ell\ell}z^k.
\]
{}From which we deduce that
\[
\begin{aligned}
\psi_{i,1}&=-\frac{a_{ni}}{\tau_0},\\
\psi_{in}&=-a_{ni}\left( u_i-\frac{1}{2\tau_0}\sum_{j=1}^n
a_{nj}^2u_j^2\right ),\\
\psi_{ik}&=a_{ki}-\frac{a_{ni}}{\tau_0}\sum_{j=1}^na_{kj}a_{nj}u_j\
\text{for }k\ne 1,n.\\
\end{aligned}
\]
Then using the formula's (\ref{5.8})
it is straightforward to check that
\[
t_1=-\frac{1}{\tau_0},\ \ t_n=\frac{\sum_{j=1}^n
a_{nj}^2u_j^2}{2\tau_0},\ \
t_k=-\frac{\sum_{j=1}^n a_{kj}a_{nj}u_j}{\tau_0},
\]
and hence that
\[
\psi_{i,1}=a_{ni}t_1,\ \ \psi_{in}=a_{ni}(t_n-u_i),\ \
\psi_{ik}=a_{ki}+a_{ni}t_k,
\]
$\eta_{\alpha,\beta}=\delta_{\alpha+\beta,n+1}$
and $t^\ell=t_{n+1-\ell}$. Assume from now on that all
$a_{ni}\ne 0$. Since $\eta_{\alpha\beta}=\delta_{\alpha+\beta,n+1}$,
the solution $F(t)$ of the WDVV equations is of the form (see \cite{Du2}):
\[
F(t)=\frac{1}{2}(t^1)^2t^n+\frac{1}{2}t^1\sum_{\alpha=2}^{n-1}
t^\alpha t^{n+1-\alpha}+f(t^2,t^3,\ldots,t^n).
\]
Since $d_n=-1$, $d_\alpha=0$ for $\alpha\ne 1,n$ and $d_F=1$,
it suffices to determine $c_{nnn}$, which is
\[
c_{nnn}=\sum_{i=1}^n \frac{a_{ni}^2(t^1-u_i)^3}
{t^n}.
\]
A straightforward calculation shows that
\[
u_i=t^1-\frac{1}{a_{ni}t^n}\left (a_{1i}-\sum_{\alpha=2}^{n-1}
\left (a_{\alpha i}t^\alpha +\frac{a_{ni}}{2} t^\alpha t^{n+1-\alpha}
\right )\right).
\]
Hence,
\[
\frac{\partial^3 f}{\partial u_n^3}= \frac{1}{(t^n)^4}\left (\sum_{i=1}^n
\frac{a_{1i}}{a_{ni}}
-\sum_{\alpha=2}^{n-1}
\left (\frac{a_{\alpha i}}{a_{ni}}t^\alpha +\frac{1}{2}
t^\alpha t^{n+1-\alpha}
\right )\right)^3.
\]
and thus
\[
F(t)=\frac{1}{2}(t^1)^2t^n+\frac{1}{2}t^1\sum_{\alpha=2}^{n-1}
t^\alpha t^{n+1-\alpha}-\frac{1}{6t^n}
\left (\sum_{i=1}^n
\frac{a_{1i}}{a_{ni}}
-\sum_{\alpha=2}^{n-1}
\left (\frac{a_{\alpha i}}{a_{ni}}t^\alpha +\frac{1}{2}
t^\alpha t^{n+1-\alpha}
\right )\right)^3.
\]
Next we give the $\xi_{ij}$'s ($\alpha\ne 1,n$):
\[
\begin{aligned}
\xi_{1j}&=a_{nj}t^ne^{zu_j},\\
\xi_{\alpha j}&=\left (a_{\alpha j}+a_{nj}t^{n+1-\alpha}\right )e^{zu_j},\\
\xi_{nj}&=\left ( a_{nj}z^{-1}+\frac{1}{t^n}\left (
a_{1j}-\sum_{\alpha=2}^{n-1}\left (
a_{\alpha j}t^\alpha+\frac{a_{nj}}{2}t^\alpha t^{n+1-\alpha}
\right )
\right )\right )e^{zu_j}.\\
\end{aligned}
\]
One easily sees that $\xi_{ij}=\frac{\partial h_j}{\partial t^i}$ with
\[
\begin{aligned}
h_j=&\frac{a_{nj}t^n}{z}e^{zu_j}\\
=&\frac{a_{nj}t^n}{z}e^{t^1-\frac{1}{a_{nj}t^n}\left
(a_{1j}-\sum_{\alpha=2}^{n-1}
\left (a_{\alpha j}t^\alpha +\frac{a_{nj}}{2} t^\alpha t^{n+1-\alpha}
\right )\right)}.
\end{aligned}
\]
To see that this are deformed flat coordinates, we determine
\[
\tilde {t}^\alpha=(-)^{\delta_{\alpha 1}}\sum_{j=1}^n
a_{n+1-\alpha,j}h_j.
\]
we find
\[
\begin{aligned}
\tilde {t}^1&=1+t^1z+O(z^2),\\
\tilde {t}^\alpha&=t^\alpha+O(z),\qquad\qquad \alpha\ne 1,n\\
\tilde {t}^n&=t^nz^{-1}+O(z^0).\\
\end{aligned}
\]

Finally we calculate the $G$-function of the Frobenius manifold.
Notice that $\tilde\tau_0(X)=\tau_0(x)=-\frac{1}{t^n}$ and that
\[
\psi_{11}\psi_{21}\cdots\psi_{n1}=\prod_{i=1}^n (a_{ni}t^n).
\]
So using theorem \ref{t5.2}, we obtain that $\gamma=\frac{n-12}{24}$
and that upto an aditive constant
\[
G(t)=\frac{12-n}{24}\log(t^n).
\]

\end{document}